\documentclass[showpacs, showkeys,amssymb,preprintnumbers,nofootinbib, superscriptaddress,eqsecnum]{revtex4-1}

\usepackage{amsmath}
\usepackage{graphicx}
\usepackage{latexsym}
\usepackage{amsfonts}
\usepackage[usenames]{color}
\usepackage{url,hyperref}
\usepackage{bm}
\usepackage{textcomp}
\newcommand{\cleqn}{\setcounter{equation}{0}}
\hypersetup{colorlinks=true, citecolor=blue, linkcolor=dred, urlcolor=blue}
\definecolor{dred}{rgb}{0.75,0.08,0}

\begin{document}

%%%%%%%%%%%%%%%%%%%%%%%%%%%%%%%%%
%
% Title page
%

\title{A New Approach to Black Hole Quasinormal Modes:\\ A Review of the Asymptotic Iteration Method}

\author{H.~T.~Cho}
\email[Email: ]{htcho@mail.tku.edu.tw}
\affiliation{Department of Physics, Tamkang University, Tamsui, Taipei, Taiwan, Republic of
China}

\author{A.~S.~Cornell}
\email[Email: ]{alan.cornell@wits.ac.za}
\affiliation{National Institute for Theoretical Physics; School of Physics, University of the Witwatersrand, Wits 2050, South Africa}

\author{Jason~Doukas}
\email[Email: ]{jasonad@yukawa.kyoto-u.ac.jp}
\affiliation{Yukawa Institute for Theoretical Physics, Kyoto University, Kyoto, 606-8502, Japan}

\author{T. -R.~Huang}
\affiliation{Department of Physics, Tamkang University, Tamsui, Taipei, Taiwan, Republic of
China}

\author{Wade~Naylor}
\email[Email: ]{naylor@phys.sci.osaka-u.ac.jp}
\affiliation{International College \& Department of Physics, Osaka University, Toyonaka, Osaka 560-0043, Japan}

\begin{abstract}
We discuss an approach to obtaining black hole quasinormal modes (QNMs) using the asymptotic iteration method (AIM), initially developed to solve second order ordinary differential equations. We introduce the standard version of this method and present an improvement more suitable for numerical implementation. We demonstrate that the AIM can be used to find radial QNMs for Schwarzschild, Reissner-Nordstr\"om (RN) and Kerr black holes in a unified way. An advantage of the AIM over the standard continued fraction method (CFM) is that for differential equations with more than three regular singular points Gaussian eliminations are not required. However, the convergence of the AIM depends on the location of the radial or angular position, choosing the best such position in general remains an open problem. This review presents for the first time the spin $0, 1/2$ \& $2$ QNMs of a Kerr black hole and the gravitational and electromagnetic QNMs of the RN black hole calculated via the AIM, and confirms results previously obtained using the CFM. We also presents some new results comparing the AIM to the WKB method. Finally we emphasize that  the AIM is well suited to higher dimensional generalizations and we give an example of doubly rotating black holes. 
\\
\end{abstract}

%\pacs{03.65.Nk; 04.70.-s}
\keywords{asymptotic iteration method, quasinormal modes, extra dimensions}
\date{18$^{th}$ November, 2011}
\preprint{YITP-11-97, WITS-CTP-83, OU-HET-735/2011}
\maketitle

\tableofcontents

\newpage
%%%%%%%%%%%%%%%%%%%%%%%%%%%%%%%%%
%
% Section 1: Introduction
%

\section{Introduction}\label{sec:1}\cleqn

\par The study of quasinormal modes (QNMs) of black holes is an old and well established subject, where the various frequencies are indicative of both the parameters of the black hole and the type of emissions possible. Initially the calculation of these frequencies was done in a purely numerical way, which requires selecting a value for the complex frequency, integrating the differential equation, and checking whether the boundary conditions are satisfied. Note that in the following we shall use the definition that QNMs are defined as solutions of the perturbed field equations with boundary conditions:
\begin{equation}
\psi(x) \to \left\{
\begin{array}{cl}
e^{-i \omega x}   &\qquad x\to -\infty \\
e^{i \omega x}  &\qquad x\to \infty
\end{array}
\right. \; , \label{QNMdef}
\end{equation}
for an $e^{-i \omega t}$ time dependence (which corresponds to ingoing waves at the horizon and outgoing waves at infinity). Also note the boundary condition as $x\to\infty$ does not apply to asymptotically anti-de Sitter spacetimes, where instead something like a Dirichlet boundary condition is imposed, for example see Ref. \cite{Moss:2001ga}. Since those conditions are not satisfied in general, the complex frequency plane must be surveyed for discrete values that lead to QNMs. This technique is time consuming and cumbersome, making it difficult to systematically survey the QNMs for a wide range of parameter values. Following early work by Vishveshwara \cite{Vishveshwara:1970zz}, Chandrasekhar and Detweiler \cite{Chandrasekhar:1975zza} pioneered this method for studying QNMs.

\par In order to improve on this, a few semi-analytic analyses were also attempted. In one approach, employed by Mashoon {\it et al.} \cite{Ferrari:1984zz}, the potential barrier in the effective one-dimensional Schr\"{o}dinger equation is replaced by a parameterized analytic potential barrier function for which simple exact solutions are known. The overall shape approximates that of the true black hole barrier, and the parameters of the barrier function are adjusted to fit the height and curvature of the true barrier at the peak. The resulting estimates for the QNM frequencies have been applied to the Schwarzschild, Reissner-Nordstr\"om and Kerr black holes, with agreement within a few percent with the numerical results of Chandrasekhar and Detweiler in the Schwarzschild case \cite{Chandrasekhar:1975zza}, and with Gunter \cite{Gunter:1980} in the Reissner-Nordstr\"om case. However, as this method relies upon a specialized barrier function, there is no systematic way to estimate the errors or to improve the accuracy.

\par The method by Leaver \cite{Leaver1985}, which is a hybrid of the analytic and the numerical, successfully generates QNM frequencies by making use of an analytic infinite-series representation of the solutions, together with a numerical solution of an equation for the QNM frequencies which involves, typically by applying a Frobenius series solution approach, the use of continued fractions. This technique is known as the continued fraction method (CFM).

\par Historically, another commonly applied technique is the WKB approximation \cite{Seidel:1989bp}. Even though it is based on an approximation, this approach is powerful as the WKB approximation is known in many cases to be more accurate, and can be carried to higher orders, either as a means to improve accuracy or as a means to estimate the errors explicitly. Also it allows a more systematic study of QNMs than has been possible using outright numerical methods. The WKB approximation has since been extended to sixth-order \cite{Konoplya:2003ii}.

\par However, all of these approaches have their limitations, where in recent years a new method has been developed which can be more efficient in some cases, called the asymptotic iteration method (AIM). Previously this method was used to solve eigenvalue problems \cite{Ciftci:2005xn} as a semi-analytic technique for solving second-order homogeneous linear differential equations. It has also been successfully shown by some of the current authors that the AIM is an efficient and accurate technique for calculating QNMs \cite{Cho:2009cj}. 

\par As such, we will review the AIM as applied to a variety of black hole spacetimes, making (where possible) comparisons with the results calculated by the WKB method and the CFM \'a la Leaver \cite{Leaver1985}. Therefore, the structure of this paper shall be: In Sec. \ref{sec:2} we shall review the AIM and the improved method of Ciftci {\it et al.} \cite{Ciftci:2005xn} (also see Ref. \cite{Barakat:2006ki}), along with a discussion of how the QNM boundary conditions are ensured. Applications to simple concrete examples, such as the harmonic oscillator and the Poschl-Teller potential are also provided. In Sec. \ref{sec:3} the case of Schwarzschild (A)dS black holes shall be discussed, developing the integer and half-spin equations. In Sec. \ref{sec:4} a review of how the QNMs of the Reissner-Nordstr\"om black holes shall be made, with several frequencies calculated in the AIM and compared with previous results. Sec. \ref{sec:5} will review the application of the AIM to Kerr black holes for spin $0, 1/2, 2$ fields. Sec. \ref{sec:6} will discuss the spin-zero QNMs for doubly rotating black holes. We then summarize and conclude in Sec. \ref{sec:7}. 

%%%%%%%%%%%%%%%%%%%%%%%%%%%%%%%%%
%
% Section 2: The Asymptotic Iteration Method
%

\section{The Asymptotic Iteration Method}\label{sec:2}\cleqn

\subsection{The Method}

\par To begin we shall now review the idea behind the AIM, where we first consider the homogeneous linear second-order differential equation for the function $\chi (x)$,
\begin{equation}
\chi'' = \lambda_{0} ( x ) \chi' + s_{0} ( x ) \chi \; , \label{eq:Chapter 4 Equation 14}
\end{equation}
where $\lambda_{0} ( x )$ and $s_{0} ( x )$ are functions in $C_{\infty} ( a , b )$. In order to find a general solution to this equation, we rely on the symmetric structure of the right-hand of Eq. (\ref{eq:Chapter 4 Equation 14}) \cite{Ciftci:2005xn}. If we differentiate Eq. (\ref{eq:Chapter 4 Equation 14}) with respect to $x$, we find that
\begin{equation*}
\chi''' = \lambda_{1} ( x ) \chi' + s_{1} ( x ) \chi \; ,
\end{equation*}
where
\begin{equation*}
\lambda_{1} = \lambda'_{0} + s_{0} + ( \lambda_{0})^{2} \; \mathrm{and} \; s_{1} = s'_{0} + s_{0} \lambda_{0} \; .
\end{equation*}
Taking the second derivative of Eq. (\ref{eq:Chapter 4 Equation 14}) we get
\begin{equation*}
\chi'''' = \lambda_{2} ( x ) \chi' + s_{2} ( x ) \chi \; ,
\end{equation*}
where
\begin{equation*}
\lambda_{2} = \lambda'_{1} + s_{1} + \lambda_{0} \lambda_{1} \hspace{1cm} \mathrm{and} \hspace{1cm} s_{1} = s'_{0} + s_{0} \lambda_{0} \; .
\end{equation*}
Iteratively, for the $( n + 1 )^{ t h }$ and the $( n + 2 )^{ th }$ derivatives, $n = 1 ,  2 , . . .$, we have
\begin{equation}
\chi^{ ( n + 1 ) } = \lambda_{n - 1} ( x ) \chi' + s_{n - 1} ( x ) \chi \; , \label{eq:Chapter 4 Equation 15}
\end{equation}
and thus bringing us to the crucial observation in the AIM is that differentiating the above equation $n$ times with respect to $x$, leaves a symmetric form for the right hand side:
\begin{equation}
\chi^{(n+2)} = \lambda_n(x) \chi' + s_n(x) \chi \; ,
\end{equation}
where
\begin{equation}
\lambda_n(x) = \lambda'_{n-1} (x)+ s_{n-1}(x) + \lambda_0(x) \lambda_{n-1}(x)
\hspace{1cm} \mathrm{and} \hspace{1cm} s_n(x) = s'_{n-1}(x) + s_0(x) \lambda_{n-1}
(x) \; . \label{eqn:2-2}
\end{equation}
For sufficiently large $n$ the asymptotic aspect of the ``method" is introduced, that is:
\begin{equation}
\frac{s_n (x)}{\lambda_n (x)} = \frac{s_{n-1}(x)}{\lambda_{n-1}(x)} \equiv \beta(x)
\; , \label{eqn:2-3}
\end{equation}
where the QNMs are obtained from the ``quantization condition"
\begin{equation}
\delta_n = s_n \lambda_{n-1} - s_{n-1} \lambda_n =0 \;, \label{eqn:quantcond}
\end{equation}
which is equivalent to imposing a termination to the number of iterations \cite{Barakat:2006ki}. From the ratio of the $(n+1)^{th}$ and the $(n+2)^{th}$ derivatives, we have
\begin{equation}
\frac{d}{dx} \ln (\chi^{(n+1)}) = \frac{\chi^{(n+2)}}{\chi^{(n+1)}} = \frac{\lambda_n \left( \chi' + \frac{s_n}{\lambda_n}\chi\right)}{\lambda_{n-1} \left( \chi' + \frac{s_{n-1}}{\lambda_{n-1}}\chi\right)} \; .
\end{equation}
From our asymptotic limit, this reduces to
\begin{equation}
\frac{d}{dx} \ln (\chi^{(n+1)}) = \frac{\lambda_n}{\lambda_{n-1}} \; ,
\end{equation}
which yields
\begin{equation}
\chi^{(n+1)}(x) = C_1 \exp \left( \int^x \frac{\lambda_n (x')}{\lambda_{n-1} (x')} dx' \right) = C_1 \lambda_{n-1} \exp \left( \int^x (\beta + \lambda_0) dx' \right) \; ,
\end{equation}
where $C_1$ is the integration constant and the right-hand side of Eq. (\ref{eqn:2-2}) and the definition of $\beta(x)$ have been used. Substituting this into Eq. (\ref{eq:Chapter 4 Equation 15}), we obtain the first-order differential equation
\begin{equation}
\chi' + \beta\chi =  C_1 \exp \left( \int^x (\beta + \lambda_0) dx' \right) \; ,
\end{equation}
which leads to the general solution
\begin{equation}
\chi(x) = \exp \left[ - \int^x \beta (x') dx' \right] \left( C_2 + C_1 \int^x \exp \left
\{ \int^{x'} \left[ \lambda_0 (x'') + 2 \beta ( x'') \right] dx'' \right\} dx' \right) \; .
\end{equation}
The integration constants, $C_1$ and $C_2$, can be determined by an appropriate choice of normalisation. Note, that for the generation of exact solutions $C_1=0$.

%%%%%%%%%%%%%%%%%%%%%%%%%%%%%%%%%

\subsection{The Improved Method}

\par Ciftci {\it et al.} \cite{Ciftci:2005xn} were among the first to note that an unappealing feature of the recursion relations in Eqs. (\ref{eqn:2-2}) is that at each iteration one must take the derivative of the $s$ and $\lambda$ terms of the previous iteration. This can slow the numerical implementation of the AIM down considerably and also lead to problems with numerical precision. 

\par To circumvent these issues we developed an improved version of the AIM which bypasses the need to take derivatives at each step \cite{Cho:2009cj}. This greatly improves both the accuracy and speed of the method. We expand the $\lambda_n$ and $s_n$ in a Taylor series around the point at which the AIM is performed, $\xi$:
\begin{eqnarray}
\lambda_n(\xi)&=&\sum_{i=0}^{\infty}c_n^i(x-\xi)^i,\\
s_n(\xi)&=&\sum_{i=0}^{\infty}d_n^i(x-\xi)^i,
\end{eqnarray}
where the $c_n^i$ and $d_n^i$ are the $i^{th}$ Taylor coefficient's of $\lambda_n(\xi)$ and $s_n(\xi)$ respectively. Substituting these expressions into Eqs. (\ref{eqn:2-2}) leads to a set of recursion relations for the coefficients:
\begin{eqnarray}
c_n^i&=&(i+1)c_{n-1}^{i+1}+d_{n-1}^{i}+\sum_{k=0}^{i}c_0^kc_{n-1}^{i-k}\; ,\label{eqn:coeffiterlam}\\
d_n^i&=&(i+1)d_{n-1}^{i+1}+\sum_{k=0}^id_0^kc_{n-1}^{i-k}\; . \label{eqn:coeffiters}
\end{eqnarray}
In terms of these coefficients the ``quantization condition'' Eq. (\ref{eqn:quantcond}) can be re-expressed as
\begin{equation}
d_n^{0} c_{n-1}^{0}-d_{n-1}^{0}c_n^0=0 \; ,
\end{equation}
and thus we have reduced the AIM into a set of recursion relations which no longer require derivative operators.

\par Observing that the right hand side of Eqs. (\ref{eqn:coeffiterlam}) and (\ref{eqn:coeffiters}) involve terms of order at most $n-1$, one can recurse these equations until only $c_0^i$ and $d_0^i$ terms remain (that is, the coefficients of $\lambda_0$ and $s_0$ only). However, for large numbers of iterations, due to the large number of terms, such expressions become impractical to compute. We avert this combinatorial problem by beginning at the $n=0$ stage and calculating the $n+1$ coefficients sequentially until the desired number of recursions is reached. Since the quantisation condition only requires the $i=0$ term, at each iteration $n$ we only need to determine coefficients with $i<N-n$, where $N$ is the maximum number of iterations to be performed. The QNMs that we calculate in this paper will be determined using this improved AIM.

%%%%%%%%%%%%%%%%%%%%%%%%%%%%%%%%%

\subsection{Two Simple Examples}

\subsubsection{The Harmonic Oscillator}\label{sec:2-1}

\par In order to understand the effectiveness of the AIM, it is appropriate to apply this method to a simple concrete problem: The harmonic oscillator potential in one dimension,
\begin{equation}
\left( - \frac{d^2}{dx^2} + x^2 \right) \phi = E \phi \; . \label{harm1}
\end{equation}
When $|x|$ approaches infinity, the wave function $\phi$ must approach zero. Asymptotically the function $\phi$ decays like a Gaussian distribution, in which case we can write
\begin{equation}
\phi (x) = e^{-x^2/2}f(x) \; , \label{harm2}
\end{equation}
where $f(x)$ is the new wave function. Substituting Eq. (\ref{harm2}) into Eq. (\ref{harm1}) then re-arranging the equation and dividing by a common factor, one can obtain
\begin{equation}
\frac{d^2 f}{dx^2} = 2 x \frac{df}{dx} + (1 - E)f \; . \label{harm3}
\end{equation}
We recognise this as Hermite's equation. For convenience we let $1 - E = - 2 j$, such that in our case $\lambda_0 = 2 x$ and $s_0 = - 2 j$. We define
\begin{equation}
\delta_n = \lambda_n s_{n-1} - \lambda_{n-1} s_n \; , \hspace{1cm} \mathrm{for} \hspace{1cm} n = 1, 2, 3, \ldots \label{harm4}
\end{equation}
Thus using Eqs. (\ref{eqn:2-2}) one can find that
\begin{equation}
\delta_n = 2^{n+1} \prod_{i=0}^n (j - i) \; , \label{harm5}
\end{equation}
and the termination condition Eq. (\ref{eqn:quantcond}) can be written as $\delta_n = 0$. Hence $j$ must be a non-negative integer, which means
\begin{equation}
E_k = 2 k +1 \; , \hspace{1cm} \mathrm{for}\hspace{1cm} k = 0 , 1, 2, \ldots \label{harm6}
\end{equation}
and this is the exact spectrum for such a potential. Moreover, the wave function $\phi(x)$ can also be derived in this method.

\par We should point out that in this case the termination condition, $\delta_n = 0$, is dependent only on the eigenvalue $j$ for a given iteration number $n$, and this is the reason why we can obtain an exact eigenvalue. However, for the black hole cases in  subsequent sections, the termination condition depends also on $x$, and therefore one can only obtain approximate eigenvalues by terminating the procedure after $n$ iterations.

%%%%%%%%%%%%%%%%%%%%%%%%%%%%%%%%%

\subsubsection{The Poschl-Teller Potential}\label{sec:2-2}

\par To conclude this section we will also demonstrate that the AIM can be applied to the case of QNMs, which have unbounded (scattering) like potentials, by recalling that we can find QNMs for Scarf II (upside-down Poschl-Teller-like) potentials \cite{OzerRoy}. This is based on observations made by one of the current authors \cite{ChoLin} relating QNMs from quasi-exactly solvable models. Indeed bound state Poschl-Teller potentials have been used for QNM approximations previously by inverting black hole potentials \cite{Ferrari:1984zz}. However, the AIM does not require any inversion of the black hole potential as we shall show.

\par Starting with the potential term
\begin{equation}
V(x) = \frac{1}{2} \mathrm{sech}^2 x \; ,
\end{equation}
and the Schr\"odinger equation, we obtain:
\begin{equation}
\frac{d^2 \psi}{dx^2} + \left( \omega^2 - \frac{1}{2}  \mathrm{sech}^2 x \right) \psi = 0 \; .
\end{equation}
As we shall also see in the following sections, it is more convenient to transform our coordinates to a finite domain. Hence, we shall use the transformation $y = \tanh x$, which leads to
\begin{eqnarray}
(1 - y^2) \frac{d}{dy} \left[ (1-y^2) \frac{d\psi}{dy} \right] + \left[ \omega^2 - \frac{1}{2}\left(1-y^2\right)\right] \psi &=& 0  \; , \nonumber \\
\Rightarrow \frac{d^2\psi}{dy^2} - \left( \frac{2 y}{1-y^2}\right) \frac{d\psi}{dy} + \left[ \frac{\omega^2}{(1-y^2)^2} - \frac{1}{2(1-y^2)} \right] \psi &=& 0 \; ,
\end{eqnarray}
where $-1<y<1$. The QNM boundary conditions in Eq. (\ref{QNMdef}) can then be implemented as follows. As $y \to 1$ we shall have $\psi \sim e^{\mp i \omega x}\sim ( 1- y)^{\pm i \omega/2} $. Hence our boundary condition $\psi \sim e^{i \omega x} \Rightarrow \psi \sim (1-y)^{- i \omega /2}$. Likewise, as $y \to -1$ we have $\psi \sim e^{\pm i \omega x}\sim (1+y)^{\pm i \omega/2} $ and the boundary condition $\psi \sim e^{- i \omega x} \Rightarrow \psi \sim (1+y)^{-i \omega/2}$. As such we can take the boundary conditions into account by writing 
\begin{equation}
\psi = (1-y)^{-i\omega/2}(1+y)^{-i \omega/2} \phi \; ,
\end{equation}
and therefore have
\begin{equation}
\frac{d^2 \phi}{dy^2} = \frac{2 y (1 - i \omega)}{1 - y^2} \frac{d\phi}{dy} + \frac{1 - 2 i \omega - 2 \omega^2}{2 ( 1-y^2)} \phi \; ,
\end{equation}
where
\begin{eqnarray}
\lambda_0 &=& \frac{2 y (1 - i \omega)}{1 - y^2} \; , \\
s_0 &=&  \frac{1 - 2 i \omega - 2 \omega^2}{2 ( 1-y^2)} \; .
\end{eqnarray}
Following the AIM procedure, that is, taking $\delta_{n}=0$ successively for $n=1,2,\cdots$, one can obtain exact eigenvalues:
\begin{equation}
\omega_{n}=\pm\frac{1}{2}-i\left(n+\frac{1}{2}\right).
\end{equation}
This exact QNM spectrum is the same as the one in Ref. \cite{ChoLin} obtained through algebraic means.

\par The reader might wonder about approximate results for cases where Poschl-Teller approximations can be used, such as Schwarzschild and SdS backgrounds, e.g., see Refs. \cite{Ferrari:1984zz, Moss:2001ga}. In fact when the black hole potential can be modeled by a Scarf like potential the AIM can be used to find the eigenvalues exactly \cite{OzerRoy} and hence the QNMs numerically. We demonstrate this in the next section. 

%%%%%%%%%%%%%%%%%%%%%%%%%%%%%%%%%
%
% Section 3: Schwarzschild (A)dS black holes
%

\section{Schwarzschild (A)dS Black Holes}\label{sec:3}\cleqn

\par We shall now begin the core focus of this review, the study of black hole QNMs using the AIM. Recall that the perturbations of the Schwarzschild black holes are described by the Regge-Wheeler \cite{Regge:1957td} and Zerilli \cite{Zerilli:1971wd} equations, and the perturbations of Kerr black holes are described by the Teukolsky equations \cite{Teukolsky:1972my}. The perturbation equations for Reissner-Nordstr\"om black holes were also derived by Zerilli \cite{Zerilli:1974ai}, and by Moncrief \cite{Moncrief:1974gw,Moncrief:1974ng,Moncrief:1975sb}. Their radial perturbation equations all have a one-dimensional Schr\"odinger-like form with an effective potential.

\par Therefore, we shall commence in the coming subsections by describing the radial perturbation equations of Schwarzschild black holes first, where our perturbed metric shall be $g_{\mu\nu} = g^0_{\mu\nu} + h_{\mu\nu}$, and where $g^0_{\mu\nu}$ is spherically symmetric. As such it is natural to introduce a mode decomposition to $h_{\mu\nu}$. Typically we write
\begin{equation}
\Psi_{lm} (t, r, \theta, \phi) = \frac{e^{-i\omega t} u_l(r)}{r} Y_{lm}(\theta,\phi) \; ,
\end{equation}
where $Y_{lm}(\theta,\phi)$ are the standard spherical harmonics. The function $u_l(r,t)$ then solves the wave equation
\begin{equation}
\left( \frac{d^2}{d x^2} - \omega^2 - V_l (r) \right) u_l (r) = 0 \; , \label{eqn:3-3}
\end{equation}
where $x$, defined by $dx = dr/f(r)$, are the so-called tortoise coordinates and $V(x)$ is a master potential of the form \cite{Berti:2009kk}
\begin{equation}
V(r) = f(r) \left[ \frac{\ell (\ell +1)}{r^2} +(1-s^2)\left( \frac{2M}{r^3} - {(4-s^2) \Lambda\over 6} \right)\right] \; . \label{mastpot}
\end{equation}
In this section
\begin{equation}
f(r)= 1- {2M\over r} -{\Lambda\over 3}r^2 \; , 
\end{equation}
with cosmological constant $\Lambda$. Here $s=0,1,2$ denotes the spin of the perturbation: scalar, electromagnetic and gravitational (for half-integer spin see Refs. \cite{Brill:1957,Medved:2003rga,Cho:2003qe} and Sec. \ref{sec:3-2}). 

%%%%%%%%%%%%%%%%%%%%%%%%%%%%%%%

\subsection{The Schwarschild Asymptotically Flat Case}\label{sec:3-11}

\par To explain the AIM we shall start with the simplest case of the radial component of a perturbation of the Schwarzschild metric outside the event horizon \cite{Zerilli:1971wd}. For an asymptotically flat Schwarzschild solution ($\Lambda=0$)
\begin{equation}
f(r)= 1- {2M\over r} \; ,
\end{equation}
where from $dx = dr/f(r)$ we have
\begin{equation}
x(r)=r + 2M \ln\left({r\over 2M}-1\right)\; ,
\end{equation}
for the tortoise coordinate $x$. 

\par Note that for the Schwarzschild background the maximum of this potential, in terms of $r$, is given by \cite{Iyer:1986nq}
\begin{equation}
r_0= {3M\over 2} {1\over \ell (\ell+1)} \Big[ \ell(\ell
+1) - (1-s^2) + \big(\ell^2(\ell+1)^2 + {14\over 9} \ell(\ell+1)(1-s^2) +(1-s^2)^2\big)^{1/2}\Big]\; .
\end{equation}

\par The choice of coordinates is somewhat arbitrary and in the next section (for SdS) we will see how an alternative choice leads to a simpler solution. Firstly, consider the change of variable:
\begin{equation}
\xi =1 - \frac{2 M}{r} \;,
\end{equation}
with $0 \leq \xi < 1$. In terms of $\xi$, our radial equation then becomes
\begin{equation}
\frac{d^2 \psi}{d\xi^2} + \frac{1 - 3 \xi}{\xi (1 - \xi)} \frac{d \psi}{d\xi} + \left[ \frac{4 M^2\omega^2}{\xi^2 (1 - \xi)^4} - \frac{\ell (\ell + 1)}{\xi (1 - \xi)^2}  -\frac{1-s^2}{\xi (1- \xi)} \right] \psi = 0 \; .
\end{equation}

\par To accommodate the out-going wave boundary condition $\psi \to e^{i\omega x}=e^{i \omega(r+2M \ln(r/2M-1))}$ as $(x,r)\to \infty$ in terms of $\xi$ (which is the limit $\xi \to  1$) and the regular singularity at the event horizon ($\xi\to 0$), we define
\begin{equation}
\psi (\xi)=  \xi^{- 2iM\omega}  (1-\xi)^{-2iM\omega} e^{\frac{2 i M \omega}{1 - \xi} }\chi (\xi) \; ,
\end{equation}
where the Coulomb power law is included in the asymptotic behaviour  (cf. Ref. \cite{Leaver1985} Eq. (5)). The radial equation then takes the form:
\begin{eqnarray}
\chi''& = & \lambda_0(\xi)  \chi' + s_0(\xi) \chi \; ,
\label{AIMform}
\end{eqnarray}
where
\begin{eqnarray}
\lambda_0(\xi) & = & \frac{4M i \omega (2\xi^2 - 4 \xi + 1) - ( 1 - 3 \xi)(1 - \xi)}{\xi (1 - \xi)
^2} \;  , \\
s_0 (\xi)& = &  \frac{16M^2  \omega^2(\xi - 2) - 8M i \omega ( 1 - \xi)+\ell (\ell + 1) +
(1-s^2)(1 - \xi)}{\xi (1 - \xi)^2} \; .
\end{eqnarray}
Note that primes of $\chi$ denote derivatives with respect to $\xi$.

\par Using these expressions we have tabulated several QNM frequencies and compared them to the WKB method of Ref. \cite{Iyer:1986nq} and the CFM of Ref. \cite{Leaver1985} in Table \ref{tab:1}. For completeness Table \ref{tab:1} also includes results from an approximate semi-analytic 3rd order WKB
method \cite{Iyer:1986nq}. More accurate semi-analytic results with better agreement to Leaver's method can be obtained by extending the WKB method to 6th order \cite{Konoplya:2003ii} and indeed in Sec. \ref{sec:5} we use this to compare with the AIM for results where the CFM has not been tabulated. 

\par It might also be worth mentioning that a different semi-analytic perturbative approach has recently been discussed by Dolan and Ottewill \cite{Dolan:2009nk}, which has the added benefit of easily being extended to any order in a perturbative scheme.

%==================================================
\begin{table}[t]
\caption{\small\sl QNMs to 4 decimal places for gravitational perturbations ($s = 2$) where the fifth column is taken from Ref. \cite{Iyer:1986nq}. Note that the imaginary part of the $n=0$, $\ell=2$ result in \cite{Iyer:1986nq} has been corrected to agree with Ref. \cite{Leaver1985}. [*] Note also that if the number of iterations in the AIM is increased, to say $50$, then we find agreement with Ref. \cite{Leaver1985} accurate to $6$ significant figures.}
\label{tab:1}
\begin{ruledtabular}
\begin{tabular}{|c|c|c|c|c|}
$\ell$ & $n$ & $\omega_{Leaver}$  & $\omega_{AIM} ~{\rm(after~15~iterations)} $ & $
\omega_{WKB}$ \\
\hline
2 & 0 & 0.3737 - 0.0896 i[*]  & 0.3737 - 0.0896 i & 0.3732 - 0.0892 i \\
&&&($<$0.01\%)($<$0.01\%)& (-0.13\%)(0.44\%)[*] \\
& 1 & 0.3467 - 0.2739 i   & 0.3467 - 0.2739 i & 0.3460 - 0.2749 i \\
&&&($<$0.01\%)($<$0.01\%)& (-0.20\%)(-0.36\%) \\
& 2 & 0.3011 - 0.4783 i  & 0.3012 - 0.4785 i & 0.3029 - 0.4711 i \\
&&&(0.03\%)(-0.04\%)& (0.60\%)(1.5\%) \\
& 3 &0.2515 - 0.7051 i & 0.2523 - 0.7023 i & 0.2475 - 0.6703 i \\
&&&(0.32\%)(0.40\%)& (-1.6\%)(4.6\%) \\
\hline
3 & 0 &0.5994 - 0.0927 i  & 0.5994 - 0.0927 i & 0.5993 - 0.0927 i \\
&&&($<$0.01\%)($<$0.01\%)& (-0.02\%)(0.0\%) \\
& 1 & 0.5826 - 0.2813 i   & 0.5826 - 0.2813 i & 0.5824 - 0.2814 i \\
&&&($<$0.01\%)($<$0.01\%)& (-0.03\%)(-0.04\%) \\
& 2 &0.5517 - 0.4791 i & 0.5517 - 0.4791 i & 0.5532 - 0.4767 i \\
&&&($<$0.01\%)($<$0.01\%)& (0.27\%)(0.50\%) \\
& 3 &0.5120 - 0.6903 i& 0.5120 - 0.6905 i & 0.5157 - 0.6774 i \\
&&&($<$0.01\%)(-0.03\%)& (0.72\%)(1.9\%) \\
& 4 &0.4702 - 0.9156 i & 0.4715 - 0.9156 i & 0.4711 - 0.8815 i \\
&&&(0.28\%)($<$0.01\%)& (0.19\%)(3.7\%) \\
& 5 &0.4314 - 1.152 i & 0.4360 - 1.147 i & 0.4189 - 1.088 i \\
&&&(1.07\%)(0.43\%)& (-2.9\%)(5.6\%) \\
\hline
4 & 0 & 0.8092 - 0.0942 i  & 0.8092 - 0.0942 i & 0.8091 - 0.0942 i \\
&&&($<$0.01\%)($<$0.01\%)& (-0.01\%)(0.0\%) \\
& 1 & 0.7966 - 0.2843 i & 0.7966 - 0.2843 i & 0.7965 - 0.2844 i \\
&&&($<$0.01\%)($<$0.01\%)& (-0.01\%)(-0.04\%) \\
& 2 & 0.7727 - 0.4799 i  & 0.7727 - 0.4799 i & 0.7736 - 0.4790 i \\
&&&($<$0.01\%)($<$0.01\%)& (0.12\%)(0.19\%) \\
& 3 &0.7398 - 0.6839 i & 0.7398 - 0.6839 i & 0.7433 - 0.6783 i \\
&&&($<$0.01\%)($<$0.01\%)& (0.47\%)(0.82\%) \\
& 4 &0.7015 - 0.8982 i& 0.7014 - 0.8985 i & 0.7072 - 0.8813 i \\
&&&(-0.01\%)(-0.03\%)& (0.81\%)(1.9\%)
\end{tabular}
\end{ruledtabular}
\end{table}
%==================================================

%%%%%%%%%%%%%%%%%%%%%%%%%%%%%%%%%

\subsection{The de-Sitter Case}\label{sec:3-12}

\par We have presented the QNMs for Schwarzchild gravitational perturbations in Table \ref{tab:1}, however, to further justify the use of this method, it is instructive to consider some more general cases. As such, we shall now consider the Schwarzschild de Sitter (SdS) case, where we have the same WKB-like wave equation and potential as in the radial equation earlier, though now
\begin{equation}
f(r) = 1 - {2M\over r} - \Lambda {r^2\over 3}\; ,
\end{equation}
where $\Lambda >0$ is the cosmological constant. Interestingly the choice of coordinates we use here leads to a simpler AIM solution, because there is no Coulomb power law tail; however, in the limit $\Lambda=0$ we recover the Schwarzschild results. Note that although it is possible to find an expression for the maximum of the potential in the radial equation, for the SdS case, it is the solution of a cubic equation, which for brevity we refrain from presenting here. In our AIM code we use a numerical routine to find the root to make the code more general.

\par In the SdS case it is more convenient to change coordinates to $\xi = 1/r$ \cite{Moss:2001ga}, which leads to the following master equation (cf. Eq. (\ref{mastpot}))
\begin{equation}
\frac{d^2 \psi}{d\xi^2} + \frac{p'}{p} \frac{d\psi} {d\xi} + \left[ \frac{\omega^2}{p^2}  - { \ell (\ell + 1)+ (1 - s^2)\left(2 M \xi - (4-s^2){\Lambda\over 6\xi^2}\right)  \over p}\right]\psi = 0\; , \label{masterxi}
\end{equation}
where we have defined
\begin{equation}
p= \xi^2 - 2M \xi^3 -\Lambda /3 \hspace{1cm} \Rightarrow \hspace{1cm} p' = 2\xi(1-3M\xi) \; .
\end{equation}
It may be worth mentioning that for SdS we can express \cite{Moss:2001ga}:
\begin{equation}
e^{i\omega x} =  (\xi-\xi_1)^{{i\omega\over2 \kappa_1}} (\xi-\xi_2)^{{i\omega\over2 \kappa_2}} (\xi-\xi_3)^{{i\omega\over2 \kappa_3}}
\end{equation}
in terms of the roots of $f(r)$, where $\xi_1$ is the event horizon and $\xi_2$ is the cosmological horizon (and $\kappa_n$ is the surface gravity at each $\xi_n$). This is useful for choosing the appropriate scaling behaviour for QNM boundary conditions.

\par Based on the above equation an appropriate choice for QNMs is to scale out the divergent behaviour at the cosmological horizon:\footnote{Note that this is opposite to the case presented in Ref.~\cite{Moss:2001ga}, where they define the QNMs as solutions with boundary conditions $\psi(x)\propto e^{\mp i \omega x}$ as $x\to \pm \infty$, for $e^{i \omega t}$ time dependence.}
\begin{equation}
\psi(\xi) = e^{i\omega x} u (\xi) \; ,  \label{SdScale}
\end{equation}
which implies
\begin{equation}
p u'' + (p'- 2 i\omega)u' - \left[\ell(\ell+1)+ (1 - s^2)\left(2 M \xi - (4-s^2){\Lambda\over 6\xi^2} \right)\right]u=0 \;  ,  \label{youeq}
\end{equation}
in terms of $\xi$. Furthermore, based on the scaling in Eq. (\ref{SdScale}), the correct  QNM condition at the horizon $\xi_1$ implies
\begin{equation}
u(x)=  (\xi-\xi_1)^{-{i\omega\over\kappa_1}}\chi(x) \; ,
\end{equation}
where 
\begin{equation}
\kappa_1 = \left.\frac 1 2 {d f\over dr}\right|_{r\to r_1} =M \xi_1^2 - \frac 1 3 {\Lambda\over \xi_1} \; ,
\end{equation}
with $\xi_1=1/r_1$, and $r_1$ is the smallest real solution of $f(r) =0$, implying $p=0$. The differential equation then takes the standard AIM form:
\begin{eqnarray}
\chi''& = & \lambda_0(\xi)  \chi' + s_0(\xi) \chi \;  ,
\end{eqnarray}
where
\begin{eqnarray}
\lambda_0(\xi) &=& -\frac 1 p \left[p'- {2i\omega \over \kappa_1(\xi-\xi_1)} - 2 i\omega\right] \;  , \\
s_0 (\xi) &= & \frac 1 p \left[\ell(\ell+1)+ (1-s^2) \left(2 M \xi - (4-s^2){\Lambda\over 6\xi^2} \right)\ +{i \omega\over \kappa_1(\xi-\xi_1)^2}\Big({i\omega\over\kappa_1} +1\Big) +(p'- 2 i\omega) {i\omega\over \kappa_1(\xi-\xi_1)}\right] \; .
\end{eqnarray}
Using these equations, we present in Table \ref{tab:2} results for SdS with $s=2$.

\par Identical results were generated by the AIM and  CFM, both after 50 iterations. Though results are presented for $n = 1, 2, 3$, $\ell = 2, 3$ modes only, the AIM is robust enough to be applied to any other case; where like the $\Lambda = 0$ case, agreement with other methods in more extreme parameter choices would only require further iterations.

\par As far as we are aware only Ref. \cite{Zhidenko:2003wq} (who used a semi-analytic WKB approach) has presented tables for  general spin fields for the SdS case. We have also compared our results to those in Ref. \cite{Zhidenko:2003wq} for the $s=0,1$ cases and find identical results (to a given accuracy in the WKB method).

\par It may  be worth mentioning that a set of three-term recurrence relations was derived in Ref. \cite{Cho:2009cj} for the CFM, valid for electromagnetic and gravitational perturbations ($s=1,2$),  while for $s=0$ this reduces to a five-term recurrence relation. However, for the AIM we can treat the $s=0,1,2$ perturbations on an equal footing, see Ref. \cite{Cho:2009cj} for more details.  Typically $n$ Gaussian Elimination steps are required to reduce an $n+3$ recurrence to a $3$-term continued fraction, e.g., for Reissner-N\"ordtrom see Ref. \cite{Leaver:1990zz} and for higher-dimensional Schwarzschild backgrounds  see Ref. \cite{Zhidenko:2006rs} (for an application of the CFM to higher dimensional asymptotic QNMs see \cite{Cardoso:2003vt}). However, all that is necessary in the AIM is to factor out the correct asymptotic behaviour at the horizon(s) and infinity (we showed this for higher-dimensional scalar spheroids in Ref. \cite{Cho:2009wf}).

%==================================================
\begin{table}[t]
\caption{\small\sl QNMs to 6 significant figures for Schwarzschild de Sitter gravitational perturbations ($s=2$) for $\ell=2$ and $\ell=3$ modes. We only present results for the AIM method, because the results are identical to those of the CFM after a given number of iterations (in this case $50$ iterations for both methods). The $n=1,2$ modes can be compared with the results in Ref. \cite{Zhidenko:2003wq} for $s=2$.}
\label{tab:2}
\begin{ruledtabular}
\begin{tabular}{|c|c|c|c|}
$\Lambda ~(\ell=2)$ & $n=1$ & $n=2$  & $n=3$ \\
\hline
0 		& 	0.373672 -  0.0889623 i	& 	0.346711 -  0.273915 i	&  	0.301050 -  0.478281 i	\\
0.02 	& 	0.338391 -  0.0817564 i	&   0.318759 -  0.249197 i	&  	0.282732 -  0.429484	 i\\
0.04 	& 	0.298895 -  0.0732967 i	&  	0.285841 -  0.221724 i	&   	0.259992 -  0.377092	 i\\
0.06		& 	0.253289 -  0.0630425 i	&	0.245742 -  0.189791 i	& 	0.230076 -  0.319157	 i\\
0.08  	& 	0.197482 -  0.0498773 i	&  	0.194115 -  0.149787 i	&   	0.187120 -  0.250257	 i\\
0.09 	& 	0.162610 -  0.0413665 i	&   0.160789 -  0.124152 i	&   0.157042 -  0.207117	 i\\
0.10 	& 	0.117916 -  0.0302105 i	& 	0.117243 -  0.0906409 i&   	0.115876 -  0.151102	 i\\
0.11 	& 	0.0372699 -  0.00961565 i	& 	0.0372493 -  0.0288470 i	&  	0.0372081 -  0.0480784 i	\\
\end{tabular}
\hspace{2.5cm}
\begin{tabular}{|c|c|c|c|}
$\Lambda ~(\ell=3)$ & $n=1$ & $n=2$  & $n=3$ \\
\hline
0 		& 	0.599443 -  0.0927030 i	& 	0.582644 -  0.281298	 i	&  	0.551685 -  0.479093	 i\\
0.02 	& 	0.543115 -  0.0844957 i	&   0.530744 -  0.255363 i 		&  	0.507015 -  0.432059 i	\\
0.04 	& 	0.480058 -  0.0751464 i	&  	0.471658 -  0.226395 i		&   	0.455011 -  0.380773	 i\\
0.06		& 	0.407175 -  0.0641396 i	&	0.402171 -  0.192807 i		& 	0.392053 -  0.322769	 i\\
0.08  	& 	0.317805 -  0.0503821 i	&  	0.315495 -  0.151249 i		&   	0.310803 -  0.252450	 i\\
0.09 	& 	0.261843 -  0.0416439 i	&   0.260572 -  0.124969 i 		&   	0.257998 -  0.208412	 i\\
0.10 	& 	0.189994 -  0.0303145 i	& 	0.189517 -  0.0909507  i		&   	0.188555 -  0.151609	 i\\
0.11 	& 	0.0600915 -  0.00961888 i &	0.0600766 -  0.0288567 i	&  	0.0600469 -  0.0480945  i\\
\end{tabular}
\end{ruledtabular}
\end{table}
%==================================================

%%%%%%%%%%%%%%%%%%%%%%%%%%%%%%%%%

\subsection{The Spin-Zero Anti-de-Sitter Case}\label{sec:3-13}

\par There are various approaches to finding QNMs for the SAdS case (an eloquent discussion is given in the appendix of Ref. \cite{Berti:2003ud}, see also \cite{Mann1999}). One approach is that of Horowitz and Hubeny \cite{Horowitz:1999jd}, which uses a series solution chosen to satisfy the SAdS QNM boundary conditions. This method can easily be applied to all perturbations ($s=0,1,2$). The other approach is to use the Frobenius method of Leaver \cite{Leaver1985}, but instead of developing a continued fraction the series must satisfy a boundary condition at infinity, such as a Dirichlet boundary condition \cite{Moss:2001ga}.

\par The AIM does not seem easy to apply to metrics where there is an asymptotically anti-de Sitter background, because for general spin, $s$, the potential at infinity is a constant and hence would include a combination of ingoing and outgoing waves, leading to a sinusoidal dependence \cite{Cardoso:2001bb}. However, for the scalar spin zero ($s=0$) case, the potential actually blows up at infinity and is effectively a bound state problem. In this case the AIM can easily be applied as we show below.

\par Let us consider the scalar wave equation in SAdS spacetime, where $\Lambda= - 3/R^2$, and $R$ is the AdS radius. The master equation takes the same form as for the graviational case, except that the potential becomes
\begin{equation}
V=\left(1-\frac{2}{r}+r^{2}\right) \left(\frac{2}{r^{3}}+2\right)=\frac{2(r-1)(r^{2}+r+2)(r^{3}+1)}{r^{4}} \; .
\end{equation}
Here for simplicity we have taken the AdS radius $R=1$, the mass of the black hole $M=1$, and the angular momentum number $l=0$. Hence the horizon radius $r_{+}=1$. Thus, with this choice we can compare with the data in Table 3.2 on page 37 of Ref. \cite{Cardoso:2003pj} (see Table \ref{tab:3} below).

\par To implement the AIM we first look at the asymptotic behavior of $\psi$. As $r\rightarrow r_{+}=1$, the potential $V$ goes to zero. In addition,
\begin{eqnarray}
\psi&\sim&e^{\pm i\left[\frac{\omega}{4}{\rm ln}(r-1)\right]}\sim (r-1)^{\pm i\omega/4}\sim \left(1-\frac{1}{r}\right)^{\pm i\omega/4} \; .
\end{eqnarray}
For QNMs we choose the ``out-going" (into the black hole) boundary condition. That is,
\begin{equation}
\psi\sim e^{-i\omega x}\sim \left(1-\frac{1}{r}\right)^{-i\omega/4} \; .
\end{equation}

\par On the other extreme of our space, $r\rightarrow\infty$, the potential goes to infinity. This is a crucial difference from the case of gravitational perturbations. In that case, the potential goes to a constant. However, in the scalar case, as $r\rightarrow\infty$,
$\psi\sim\left(1/r\right)^{\pm\sqrt{2}+1/2}$
and to implement the Dirichlet boundary condition, we take
\begin{equation}
\psi\sim\left(\frac{1}{r}\right)^{\frac{1}{2}+\sqrt{2}} \; .
\end{equation}

\par For the AIM one possible choice of variables is
\begin{equation}
\xi=1-\frac{1}{r} \; ,
\end{equation}
and we see that to accommodate the asymptotic behaviour of the wavefunction we should take
\begin{equation}
\psi=\xi^{-i\omega/4}(1-\xi)^{\sqrt{2}+\frac{1}{2}}\chi \; .
\end{equation}
Finally, after some work we find the scalar perturbation equation is
\begin{eqnarray}
\chi''& = & \lambda_0(\xi)  \chi' + s_0(\xi) \chi \; ,
\end{eqnarray}
where
\begin{eqnarray}
\lambda_{0}&=&-\frac{-i\omega q+2[-4+2(9+4\sqrt{2})\xi-(21+10\sqrt{2})\xi^{2}+4(2+\sqrt{2})\xi^{3}]}
{2\xi q} \; , \nonumber\\ \\
s_{0}&=&\frac{1}{16\xi q^{2}}\bigg\{4i\omega[9+8\sqrt{2}-2(7+5\sqrt{2})\xi+(6+4\sqrt{2})\xi^{2}]q +\omega^{2}(-1+\xi)^{2}(-40+41\xi-20\xi^{2}+4\xi^{3})\nonumber\\
&&\ \ -4[4-5\xi+2\xi^{2}][-8(3+2\sqrt{2})+8(10+7\sqrt{2})\xi-(91+64\sqrt{2})\xi^{2}+(34+24\sqrt{2})\xi^{3}]\bigg\} \; ,\nonumber\\
\end{eqnarray}
and $q=(-4+9\xi-7\xi^{2}+2\xi^{3})$. Using the AIM we find the results presented in Table \ref{tab:3} below.

%==================================================
\begin{table}[ht]
\caption{\small\sl Comparison of the first few QNMs to 6 significant figures for Schwarzschild anti de Sitter scalar perturbations ($s=0$) for $\ell=0$ modes with $r_+=1$. The second column corresponds to data \cite{Cardoso:2003pj} using the Horowitz and Hubeny (HH) method \cite{Horowitz:1999jd}, while the third column is for the AIM using 70 iterations. [*] Note the mismatch for the real part of the $n=3$ mode in Ref. \cite{Cardoso:2003pj}; we have confirmed this using the {\it Mathematica} notebook provided in Ref. \cite{Berti:2009kk}.}
\label{tab:3}
\begin{ruledtabular}
\begin{tabular}{|c|c|c|}
$n$ & HH method & AIM \\
\hline
0	& 	2.7982 - 2.6712 i	& 		2.79823 -2.67121 i  	\\
1	& 	4.75849 - 5.03757 i	& 		4.75850 -5.03757 i  	\\
2	& 	6.71927 - 7.39449 i	& 		6.71931 -7.39450  i 	\\
3	&   8.68223[*] - 9.74852 i	&   8.68233 -9.74854  i 	\\
4	& 	10.6467 - 12.1012 i	& 		10.6469 -12.1013 i  	\\
5	& 	12.6121 - 14.4533 i	& 		12.6125 -14.4533 i  	\\
6	& 	14.5782 - 16.8049 i	& 		14.5788 -16.8050 i  	\\
7	& 	16.5449 - 19.1562 i	& 		16.5457 -19.1563 i  	
\end{tabular}
\end{ruledtabular}
\end{table}
%==================================================

%%%%%%%%%%%%%%%%%%%%%%%%%%%%%%%%%
%
% Section 4: Reissner-Nordstr\"om black holes
%

\section{Reissner-Nordstr\"om Black Holes}\label{sec:4}\cleqn

\par The procedure for obtaining the quasinormal frequencies of Reissner-Nordstr\"om black holes in four-dimensional spacetime is similar to that of our earlier cases. Starting with the Reissner-Nordstr\"om metric
\begin{equation}
ds^2 = - f(r) dt^2 + \frac{dr^2}{f(r)} + r^2 d\theta^2 + r^2 \sin^2 \theta d\phi^2 \; , \label{eqn:4-1}
\end{equation}
where $f(r) = \left( 1 - \frac{1}{r} + \frac{Q^2}{r^2} \right)$ and $|Q| \leq \frac{1}{2}$ is the charge of the black hole. If we consider perturbations exterior to the event horizon, the perturbation equations of the Reissner-Nordstr\"om (charged and non-rotating) geometry can be separated into two pairs of Schr\"odinger-like equations, which describe the even- and odd-parity oscillations respectively \cite{Zerilli:1974ai,Moncrief:1974gw,Moncrief:1974ng,Moncrief:1975sb}. They are given by
\begin{equation}
\left( \frac{d^2}{d x^2} - \rho^2 - V_i^{(\pm)} \right) Z_i^{(\pm)}=0 \; , \label{eqn:4-2}
\end{equation}
where $(+)$ corresponds to even- and $(-)$ to odd-parity modes:
\begin{eqnarray}
V_i^{(-)}(r) & = & \frac{\Delta}{r^5} \left( A r - q_j + \frac{4 Q^2}{r} \right) \; , \label{eqn:4-3} \\
V_i^{(+)}(r) & = & V_i^{(-)} (r) + 2 q_j \frac{d}{d x} \left( \frac{\Delta}{r^2 [ (l-1)(l+2) r + q_j]} \right) \; , \label{eqn:4-4}
\end{eqnarray}
for $i=j=1,2$ ($i\neq j$) where
\begin{eqnarray}
\frac{dr}{dx} & = & \frac{\Delta}{r} \; , \label{eqn:4-5} \\
\Delta & = & r^2 - r + Q^2 \equiv (r- r_+)(r-r_-) \; , \label{eqn:4-6} \\
A & = & l(l+1) \; , \label{eqn:3-7} \\
q_1 & = & \frac{1}{2} \left[ 3 + \sqrt{9 + 16 Q^2 (l - 1)(l+2)} \right] \; , \label{eqn:4-8} \\
q_2 & = & \frac{1}{2} \left[ 3 - \sqrt{9 + 16 Q^2 (l - 1)(l+2)} \right] \; , \label{eqn:4-9}
\end{eqnarray}
and $\rho = - i \omega$. Here $\omega$ is the frequency, $l$ the angular momentum parameter and $r_-$ and $r_+$ the radii of the inner and outer (event) horizons of the black hole respectively. Note that $r_+ = 1$ and $r_- = 0$ at the Schwarzschild limit $(Q=0)$; $r_+ = r_- = \frac{1}{2}$ at the extremal limit $(Q=\frac{1}{2})$. Here the tortoise coordinate is given by
\begin{equation}
x= \int \frac{r^2}{\Delta} dr = r+ \frac{r_+^2}{r_+ - r_-} \ln (r-r_+) - \frac{r^2}{r_+ - r_-} \ln (r- r_-) \; , \label{eqn:4-10}
\end{equation}
which ranges from $-\infty$ at the event horizon to $+\infty$ at spatial infinity.

\par The QNMs of the Reissner-Nordstr\"om black holes are ordinarily accompanied by the emission of both electromagnetic and gravitational radiation, except at the Schwarzschild limit \cite{Kokkotas:1988fm,Leaver:1990zz}. Eq. (\ref{eqn:4-2}) corresponds to purely gravitational perturbations for the radial wave functions $Z_2^{(\pm)}$ and purely electromagnetic perturbations for $Z_1^{(\pm)}$ at the Schwarzschild limit. Chandrasekhar \cite{Chandrasekhar:1985kt} has shown that the solution $Z_i^{(+)}$ for the even-parity oscillations and $Z_i^{(-)}$ for the odd-parity oscillations have the relationship
\begin{eqnarray}
\left[ A(A-2) - 2 \rho q_j\right]Z_i^{(+)} & = & \left\{ A(A-2) + \frac{2 q_j^2 \Delta}{r^3\left[ (A-2)r + q_j\right]} \right\} Z_i^{(-)} 
 + 2 q_j \frac{d Z_i^{(-)}}{dx} \; , \label{eqn:4-10b}
\end{eqnarray}
so one can just consider solutions for a specific parity, as in the Schwarzschild case, to understand the property of the black hole. Since the formalism of the effective potential $V_i^{(-)}$ in the odd-parity equation is much simpler than $V_i^{(+)}$ in the even-parity equation, it is customary to compute the QNMs for the odd-parity modes.

\par Note that the mass $M$ of the Reissner-Nordstr\"om black hole has been scaled to $2 M = 1$, so its quasinormal frequencies are uniquely determined by the charge $Q$, the angular momentum $l$, and the overtone number $n$ of the mode.

\par The following procedure is similar to that in Sec. \ref{sec:3}. At first we change $r$ to the variable $x$ in Eq. (\ref{eqn:4-2}) for the odd-parity mode. From Eq. (\ref{eqn:4-5}) we have
\begin{equation}
\frac{d}{dx} = \frac{\Delta}{r^2} \frac{d}{dr} \; , \label{eqn:4-11}
\end{equation}
and
\begin{equation}
\frac{d^2}{dx^2} = \left(\frac{\Delta}{r^2}\right)\left( \frac{r - 2 Q^2}{r^3} \right) \frac{d}{dr} + \left(\frac{\Delta}{r^2}\right)^2 \frac{d^2}{dr^2} \; . \label{eqn:4-12}
\end{equation}
Substituting Eq. (\ref{eqn:4-12}) into Eq. (\ref{eqn:4-2}) for the odd-parity mode, we get
\begin{equation}
\left(\frac{\Delta}{r^2}\right)^2 \frac{d^2}{dr^2} Z_i^{(-)}+ \left(\frac{\Delta}{r^2}\right)\left( \frac{r - 2 Q^2}{r^3} \right) \frac{d}{dr} Z_i^{(-)} - \left[ \rho^2 + V_i^{(-)} \right] Z_i^{(-)}= 0 \; . \label{eqn:4-13}
\end{equation}
Considering the QNM boundary conditions in the Reissner-Nordstr\"om case
\begin{equation}
Z_i^{(-)} \to \left\{ \begin{array}{cl} e^{-i \omega x} & ; x \to - \infty \\ e^{i \omega x} & ; x \to + \infty \end{array} \right. \; , \label{eqn:4-14}
\end{equation}
and incorporating this into the radial wave function $Z_i^{(-)}$, we have a form involving the asymptotic behaviour \cite{Leaver:1990zz}
\begin{equation}
Z_i^{(-)} = e^{-\rho r} r^{-1} (r-r_-)^{1 - \rho - \frac{\rho r_+^2}{r_+ - r_-}} (r - r_+)^{\frac{\rho r_+^2}{r_+ - r_-}} \chi_{Z_i} (r) \; . \label{eqn:4-15}
\end{equation}
Differentiating Eq. (\ref{eqn:4-15}) one and two times with respect to $r$, we have
\begin{equation}
Z_{i,r}^{(-)} \equiv \frac{d}{dr} Z_i^{(-)} = e^{-\rho r} r^{-1} (r-r_-)^{1 - \rho - \frac{\rho r_+^2}{r_+ - r_-}} (r - r_+)^{\frac{\rho r_+^2}{r_+ - r_-}} \left( \chi_{Z_i , r} + \Gamma_Z \chi_{Z_i} \right) \; , \label{eqn:4-16}
\end{equation}
and
\begin{equation}
Z_{i,rr}^{(-)} = e^{-\rho r} r^{-1} (r-r_-)^{1 - \rho - \frac{\rho r_+^2}{r_+ - r_-}} (r - r_+)^{\frac{\rho r_+^2}{r_+ - r_-}} \left[ \chi_{Z_i , rr} + 2 \Gamma_Z \chi_{Z_i, r} + \left( \Gamma_Z^2 + \Gamma_{Z,r} \right) \chi_{Z_i} \right] \; , \label{eqn:4-17}
\end{equation}
where $\Gamma_Z$ is defined by
\begin{equation}
\Gamma_Z = - \rho - \frac{1}{r} + \frac{(1-\rho)( r_+ - r_-)-\rho r_+^2}{(r_+ - r_-)(r - r_-)} + \frac{\rho r_+^2}{ (r_+ - r_-)(r - r_+)} \; . \label{eqn:4-18}
\end{equation}
Substituting Eqs. (\ref{eqn:4-16}) and (\ref{eqn:4-17}) into Eq. (\ref{eqn:4-13}), we obtain
\begin{eqnarray}
&&\left(\frac{\Delta}{r^2}\right)^2 \chi_{Z_i , rr} + \left[ 2 \Gamma_Z \left(\frac{\Delta}{r^2}\right)^2 + \left(\frac{\Delta}{r^2}\right) \left( \frac{r - 2 Q^2}{r^3} \right) \right] \chi_{Z_i , r} \nonumber\\
&&\ \ \ + \left\{ \left(\frac{\Delta}{r^2}\right)^2 \left( \Gamma_Z^2 + \Gamma_{Z,r} \right)
+ \left(\frac{\Delta}{r^2}\right) \left( \frac{r - 2 Q^2}{r^3} \right)\Gamma_Z - \left[ \rho^2 + V_i^{(-)} \right] \right\} \chi_{Z_i}=0 \; . \label{eqn:4-19}
\end{eqnarray}

\par For the same reason as in Sec. \ref{sec:3}, here we change the variable $r$ to $\xi$ by the definition $\xi = 1 - \frac{r_+}{r}$, which ranges from 0 at the event horizon to 1 at spatial infinity. Thus we have
\begin{equation}
\frac{d}{dr} = \frac{(1-\xi)^2}{r_+} \frac{d}{d\xi} \; , \label{eqn:4-20}
\end{equation}
and
\begin{equation}
\frac{d^2}{dr^2} = \frac{(1-\xi)^4}{r_+^2} \frac{d^2}{d\xi^2} - 2\frac{(1-\xi)^3}{r_+^2} \frac{d}{d\xi} \; . \label{eqn:4-21}
\end{equation}
Substituting Eqs. (\ref{eqn:4-20}) and (\ref{eqn:4-21}) into Eq. (\ref{eqn:4-19}), and rewriting the equation in the AIM form, we obtain
\begin{equation}
\chi_{Z_i, \xi\xi} = \lambda_{Z_i} (\xi) \chi_{Z_i , \xi} + s_{Z_i} (\xi) \chi_{Z_i} \; , \label{eqn:4-23}
\end{equation}
where
\begin{eqnarray}
\lambda_{Z_i} (\xi) & = & \frac{2}{1-\xi} - \frac{ 2 r_+ \Gamma_Z}{(1-\xi)^2} - \frac{r_+ - 2 Q^2 (1-\xi)}{\Delta (1-\xi)^2} \; , \label{eqn:4-24} \\
s_{Z_i} (\xi) & = & \frac{r_+^6 \left[ \rho^2 + V_i^{(-)}\right]}{\Delta^2 (1-\xi)^8} - \frac{r_+ \Gamma_Z}{\Delta (1- \xi)^4} \left[ r_+ - 2 Q^2 (1- \xi)\right] - \frac{r_+^2 \Gamma_Z^2}{\Delta (1-\xi)^4} - \frac{r_+ \Gamma_{Z , \xi}}{(1-\xi)^2} \; , \label{eqn:4-25} \\
\Gamma_Z & = & -\rho - \frac{1-\xi}{r_+} + \frac{\left[ (1-\rho)(r_+ - r_-)- \rho r_+^2 \right] (1-\xi)}{(r_+ - r_-)\left[ r_+ - r_-(1-\xi)\right]} + \frac{\rho r_+(1-\xi)}{(r_+ - r_-)\xi} \; , \label{eqn:4-26} \\
V_i^{(-)} & = & \Delta \frac{(1-\xi)^5}{r_+^5} \left[ \frac{A r_+}{1-\xi} - q_j + \frac{4 Q^2 (1-\xi)}{r_+}  \right] \; , \label{eqn:4-27} \\
\Delta & = & \frac{r_+ \xi \left[ r_+ - r_- (1 - \xi) \right]}{(1-\xi)^2} \; . \label{eqn:4-28}
\end{eqnarray}

\par The numerical results to four decimal places are presented in Table \ref{tab:RS3}, \ref{tab:RS4} and \ref{tab:RS5}. They are compared with $\rho_{Leaver}$ and $\rho_{WKB}$ from Refs. \cite{Leaver:1990zz} and \cite{Kokkotas:1988fm} respectively. The quasinormal frequencies appear as complex conjugate pairs in $\rho$; we list only the ones with $Im(\rho)>0$. Note that we arrange $\rho$ as $(Im(\rho),Re(\rho))$. In Table \ref{tab:RS5} the quasinormal frequencies obtained by the WKB method are not available. It is apparent that the quasinormal frequencies obtained by the AIM are very accurate except for $n = 2$ in the extremal case $Q = \frac{1}{2}$ in Tables \ref{tab:RS3} and \ref{tab:RS5}.

\par The QNMs of $l = 2$ and $i = 2$ in Table \ref{tab:RS3} reduce to the purely gravitational QNMs in the Schwarzschild case at $Q = 0$, while the QNMs of $l = 2$ and $i = 1$ in Table \ref{tab:RS4} reduce to the purely electromagnetic QNMs at $Q = 0$.

\par Some comments on the higher ($n=2$) overtones for the Reissner-Nordstr\"om black hole for the extremal limit ($Q=1/2$) are perhaps necessary. In general, much like the CFM the AIM begins to break down for larger overtones, requiring more iterations. However, near the extremal limit ($Q=1/2$) the horizons become degenerate and the singularity structure of the corresponding differential (radial) equation changes \cite{Andersson:1996xw} (the number of singular points are different in the non-extremal and the extremal cases), and causes the current implementation of the AIM (cf. \ref{eqn:4-15}) to break down. Thus, we see in Tables \ref{tab:RS3}-\ref{tab:RS5} that for $Q=0.495$ some of the values have large errors when compared to the CFM.

%==================================================
\begin{table}[ht]
\caption{\small\sl Reissner-Nordstr\"om quasinormal frequency parameter values ($\rho = -i \omega$) for the fundamental ($n = 0$) and two lowest overtones for $l = 2$ and $i = 2$.}
\label{tab:RS3}
\begin{ruledtabular}
\begin{tabular}{|c|c|c|c|c|}
$n$ & $Q$ & $\rho_{Leaver}$ & $\rho_{AIM}$ & $\rho_{WKB}$ \\
\hline
0 & 0 & (0.7473,-0.1779) & (0.7473,-0.1779)  & (0.7463,-0.1784) \\
&&& ($<$0.01\%)($<$0.01\%) & (-0.13\%)(-0.28\%) \\
0 & 0.2 & (0.7569,-0.1788) & (0.7569,-0.1788) & (0.7558,-0.1793) \\
&&& ($<$0.01\%)($<$0.01\%) & (-0.15\%)(-0.28\%) \\
0 & 0.4 & (0.8024,-0.1793) & (0.8024,-0.1793) & (0.8011,-0.1797) \\
&&& ($<$0.01\%)($<$0.01\%) & (-0.16\%)(-0.22\%) \\
0 & 0.495 & (0.8586,-0.1685) & (0.8586,-0.1685) & (0.8566,-0.1706) \\
&&& ($<$0.01\%)($<$0.01\%) & (-0.23\%)(-1.25\%) \\
\hline
1 & 0 & (0.6934,-0.5478) & (0.6934,-0.5478) & (0.6920,-0.5478) \\
&&& ($<$0.01\%)($<$0.01\%) & (-0.20\%)(-0.37\%) \\
1 & 0.2 & (0.7035,-0.5503) & (0.7035,-0.5502) & (0.7020,-0.5522) \\
&&& ($<$0.01\%)(0.02\%) & (-0.21\%)(-0.36\%) \\
1 & 0.4 & (0.7538,-0.5499) & (0.7538,-0.5499) & (0.7510,-0.5525) \\
&&& ($<$0.01\%)($<$0.01\%) & (-0.37\%)(-0.47\%) \\
1 & 0.495 & (0.8070,-0.5140) & (0.8067,-0.5164) & (0.8068,-0.5287) \\
&&& (-0.04\%)(0.47\%) & (0.01\%)(-2.21\%) \\
\hline
2 & 0 & (0.6021,-0.9566) & (0.6021,-0.9566) & (0.6059,-0.9421) \\
&&& ($<$0.01\%)($<$0.01\%) & (0.63\%)(1.52\%) \\
2 & 0.2 & (0.6129,-0.9599) & (0.6128,-0.9599) & (0.6164,-0.9458) \\
&&& (0.02\%)($<$0.01\%) & (0.57\%)(1.47\%) \\
2 & 0.4 & (0.6703,-0.9531) & (0.6703,-0.9531) & (0.6717,-0.9455) \\
&&& ($<$0.01\%)($<$0.01\%) & (0.21\%)(0.80\%) \\
2 & 0.495 & (0.7078,-0.8872) & (0.8350,-0.8347) & (0.7344,-0.9135) \\
&&& (17.97\%)(5.92\%) & (2.66\%)(-2.96\%)
\end{tabular}
\end{ruledtabular}
\end{table}
%==================================================
%==================================================
\begin{table}[ht]
\caption{\small\sl Reissner-Nordstr\"om quasinormal frequency parameter values ($\rho = -i \omega$) for the fundamental ($n = 1$) and two lowest overtones for $l = 2$ and $i = 1$.}
\label{tab:RS4}
\begin{ruledtabular}
\begin{tabular}{|c|c|c|c|c|}
$n$ & $Q$ & $\rho_{Leaver}$ & $\rho_{AIM}$ & $\rho_{WKB}$ \\
\hline
0 & 0 & (0.9152,-0.1900) & (0.9152,-0.1900)  & (0.9143,-0.1901) \\
&&& ($<$0.01\%)($<$0.01\%) & (-0.10\%)(-0.05\%) \\
0 & 0.2 & (0.9599,-0.1929) & (0.9599,-0.1929) & (0.9590,-0.1930) \\
&&& ($<$0.01\%)($<$0.01\%) & (-0.09\%)(-0.05\%) \\
0 & 0.4 & (1.1403,-0.1984) & (1.1403,-0.1981) & (1.1395,-0.1980) \\
&&& ($<$0.01\%)(0.15\%) & (-0.07\%)(-0.20\%) \\
0 & 0.495 & (1.3855,-0.1773) & (1.3855,-0.1773) & (1.3850,-0.1783) \\
&&& ($<$0.01\%)($<$0.01\%) & (-0.04\%)(-0.56\%) \\
\hline
1 & 0 & (0.8731,-0.5814) & (0.8731,-0.5814) & (0.8717,-0.5819) \\
&&& ($<$0.01\%)($<$0.01\%) & (-0.16\%)(-0.09\%) \\
1 & 0.2 & (0.9200,-0.5894) & (0.9200,-0.5894) & (0.9186,-0.5897) \\
&&& ($<$0.01\%)($<$0.01\%) & (-0.15\%)(-0.05\%) \\
1 & 0.4 & (1.1100,-0.6021) & (1.1100,-0.6021) & (1.1081,-0.6014) \\
&&& ($<$0.01\%)($<$0.01\%) & (-0.17\%)(0.12\%) \\
1 & 0.495 & (1.3573,-0.5350) & (1.3573,-0.5350) & (1.3579,-0.5423) \\
&&& ($<$0.01\%)($<$0.01\%) & (0.04\%)(-1.36\%) \\
\hline
2 & 0 & (0.8024,-1.0032) & (0.8024,-1.0032) & (0.8046,-0.9917) \\
&&& ($<$0.01\%)($<$0.01\%) & (0.27\%)(1.15\%) \\
2 & 0.2 & (0.8530,-1.0143) & (0.8530,-1.0143) & (0.8548,-1.0037) \\
&&& ($<$0.01\%)($<$0.01\%) & (0.21\%)(1.05\%) \\
2 & 0.4 & (1.0582,-1.0263) & (1.0582,-1.0263) & (1.0568,-1.0181) \\
&&& ($<$0.01\%)($<$0.01\%) & (-0.13\%)(0.80\%) \\
2 & 0.495 & (1.3019,-0.9024) & (1.3019,-0.9024) & (1.3141,-0.9222) \\
&&& ($<$0.01\%)($<$0.01\%) & (0.94\%)(-2.19\%)
\end{tabular}
\end{ruledtabular}
\end{table}
%==================================================
%==================================================
\begin{table}[ht]
\caption{\small\sl Reissner-Nordstr\"om quasinormal frequency parameter values ($\rho = -i \omega$) for the fundamental ($n = 1$) and two lowest overtones for $l = 1$ and $i = 1$.}
\label{tab:RS5}
\begin{ruledtabular}
\begin{tabular}{|c|c|c|c|}
$n$ & $Q$ & $\rho_{Leaver}$ & $\rho_{AIM}$ \\
\hline
0 & 0 & (0.4965,-0.1850) & (0.4965,-0.1850) \\
&&& ($<$0.01\%)($<$0.01\%) \\
0 & 0.2 & (0.5238,-0.1883) & (0.5238,-0.1883) \\
&&& ($<$0.01\%)($<$0.01\%) \\
0 & 0.4 & (0.6470,-0.1965) & (0.6470,-0.1965) \\
&&& ($<$0.01\%)($<$0.01\%) \\
0 & 0.495 & (0.8428,-0.1742) & (0.8428,-0.1742)\\
&&& ($<$0.01\%)($<$0.01\%) \\
\hline
1 & 0 & (0.4290,-0.5873) & (0.4292,-0.5873) \\
&&& ($<$0.01\%)(0.05\%) \\
1 & 0.2 & (0.4598,-0.5953) & (0.4598,-0.5953) \\
&&& ($<$0.01\%)($<$0.01\%) \\
1 & 0.4 & (0.5980,-0.6107) & (0.5980,-0.6107) \\
&&& ($<$0.01\%)($<$0.01\%) \\
1 & 0.495 & (0.7979,-0.5293) & (0.7978,-0.5280) \\
&&& (0.01\%)(0.25\%) \\
\hline
2 & 0 & (0.3496,-1.0504) & (0.3495,-1.0504) \\
&&& (0.03\%)($<$0.01\%) \\
2 & 0.2 & (0.3832,-1.0596) & (0.3832,-1.0596) \\
&&& ($<$0.01\%)($<$0.01\%) \\
2 & 0.4 & (0.5340,-1.0660) & (0.5340,-1.0659) \\
&&& ($<$0.01\%)($<$0.01\%) \\
2 & 0.495 & (0.7104,-0.9055) & (0.6248,-1.0574) \\
&&& (-12.05\%)(-16.78\%)
\end{tabular}
\end{ruledtabular}
\end{table}
%==================================================

%%%%%%%%%%%%%%%%%%%%%%%%%%%%%%%%%
%
% Section 5: Kerr (A)dS black holes: Change to four dimensional black holes
%

\section{Kerr Black Holes}\label{sec:5}\cleqn

\par A rotating black hole carrying angular momentum is described by the Kerr metric (in Boyer-Lindquist coordinates) as
\begin{equation}
ds^{2}=-\left(1-\frac{r}{\Sigma}\right)dt^{2}-\frac{2ar\sin^{2}\theta}{\Sigma}dtd\phi+\frac{\Sigma}{\Delta}dr^{2}
\Sigma d\theta^{2}+\left(r^{2}+a^{2}+\frac{a^{2}r\sin^{2}\theta}{\Sigma}\right)\sin^{2}\theta d\phi \; ,
\end{equation}
with 
\begin{eqnarray}
\Delta&=&r^{2}+a^2-2Mr\equiv(r-r_{-})(r-r_{+}) \; ,\\
\Sigma&=&r^{2}+a^{2}\cos^{2}\theta \; ,
\end{eqnarray}
and where $a$ is the Kerr rotation parameter with $0\leq a \leq M$, $M$ being included as a general black hole mass. The horizons $r_{-}$ and $r_{+}$ are again the inner and the outer (event) horizons respectively. Teukolsky \cite{Teukolsky:1972my} showed that the perturbation equations in the Kerr geometry are separable, where the separated equations for the angular wave function ${}_sS_{lm}(\theta)$ and the radial wave function $R(r)$ are given by:
\begin{eqnarray}
[(1-u^{2})S_{,u}]_{,u}+\left[a^{2}\omega^{2}u^{2}-2a\omega su+s+{}_sA_{lm}-\frac{(m+su)^{2}}{1-u^{2}}\right]{}_sS_{lm}&=&0 \; , \label{kerrang}\\
\Delta R_{,rr}+(s+1)(2r-1)R_{,r}+K(r)R&=&0 \; , \label{kerrrad}
\end{eqnarray}
where the function
\begin{equation}
K(r)=\frac{1}{\Delta}\left\{\left(r^{2}+a^{2}\right)^{2}\omega^{2}-2am\omega r+a^{2}m^{2}
+is\left[am\left(2r-1\right)-\omega\left(r^{2}-a^{2}\right)\right]\right\}+2is\omega r-a^{2}\omega^{2}-{}_sA_{lm} \; .
\end{equation}
In the above $u=\cos\theta$, $s$ is the spin weight, ${}_sA_{lm}$ is the spin-weighted separation constant for the angular equation, and $m$ is another angular momentum parameter. For completeness the evaluation of the separation constant ${}_sA_{lm}$ using the AIM is discussed in Appendix \ref{sec:A}.

\par In order to use the AIM we need to solve for the angular solution in the radial equation. However, for nonzero $s$ the effective potential of the radial equation is in general complex. A straight forward application of the AIM does not give the correct answer. In fact a similar problem occurs in both numerical \cite{Detweiler1977} and WKB \cite{Seidel:1989bp} methods. For this reason we shall look at each of the spin cases ($0,\frac 1 2,2$) separately in the following subsections.

%%%%%%%%%%%%%%%%%%%%%%%%%%%%%%%%%

\subsection{The Spin-Zero Case}\label{sec:5-2}

\par Because the AIM works better on a compact domain, we define a new variable $y = 1 - \frac { r _ { + } } { r }$, which ranges from $0$ at the event horizon $( r = r _ { + } )$ to $1$ at spatial infinity. It is then necessary to incorporate the boundary conditions, which expressed in the new compact domain, where
\begin{equation}
R(r)=(r^2+a^2)^{-1/2} \psi(r)
\end{equation}
 is
\begin{equation}
\psi ( y ) = \left ( 1 - \frac { r _ { - } } { r _ { + } } ( 1 - y ) \right ) ^ { - i \sigma _ { - } } y ^ { - \sigma _ { + } } ( 1 - y ) ^ { - r _ { + } \omega } e ^ { i \omega \frac { r _ { + } } { 1 - y } } \chi ( y ) \; . \label{eqn:5-6}
\end{equation}
By making the change of coordinates and change of function, Eq. (\ref{eqn:5-6}) takes the form
\begin{equation}
\chi ( y ) = \lambda _ { 0 } ( y ) + s _ { 0 } ( y ) \; ,\label{eqn:5-7}
\end{equation}
where
\begin{equation}
\lambda _ { 0 } = - 2 \frac { 1 } { g } \frac { \mathrm { d } g } { \mathrm { d } y } - \frac { 1 } { f } \frac { \mathrm { d } f } { \mathrm { d } y } \; ,
\end{equation}
and
\begin{equation}
s _ { 0 } = - \frac { 1 } { g } \frac { \mathrm { d } ^ { 2 } g } { \mathrm { d } y ^ { 2 } } - \frac { 1 } { f } \frac { \mathrm { d } f } { \mathrm { d } y } \times \frac { 1 } { g } \frac { \mathrm { d } g } { \mathrm { d } y } - \frac { 1 } { f ^ { 2 } } \left ( \omega ^ { 2 } - V | _ { r = r _ { + } ( 1 - y ) ^ { - 1 } } \right ) \; .
\end{equation}
In the above we have defined
\begin{equation}
f = \left.  \left( \frac { \Delta } { r ^ { 2 } + a ^ { 2 } } \frac { \mathrm { d } y } { \mathrm { d } r } \right) \right|_{ r = r _ { + } ( 1 - y ) ^ { - 1 } } \; ,
\end{equation}
and
\begin{equation}
g = ( 1 - y ) ^ { - 2 i \omega } \left ( 1 - \frac { r _ { - } } { r _ { + } } ( 1 - y ) \right ) ^ { i \sigma _ { - } } y ^ { - i \sigma _ { + } } e ^ { i \omega r _ { + } ( 1 - y ) ^ { - 1 } } \; ,
\end{equation}
where $\Delta = r^2 + a^2 - 2 M r$, 
\begin{equation}
\sigma _ { \pm } = \frac { 1 } { r _ { + } - r _ { - } } [ ( r ^ { 2 } _ { \pm } + a ^ { 2 } ) \omega + a m ] \; , \label{eqn:2-last}
\end{equation}
and $a$ is again our rotation parameter. Eq. (\ref{eqn:5-7}) is now in the correct form to use the AIM for QNM frequency calculations. Note that the potential, $V$ is
\begin{equation}
V=-{1\over\Delta} ((K-2(r^2+a^2))^2-\Delta\lambda)~,
\end{equation}
where the angular separation constant is defined via $\lambda=A_{l,m}+a^2\omega^2 - 2a m \omega$.\footnote{Even though the radial and angular equations are coupled via the separation constant, ${}_0A_{l,m}$, we are able to find excellent agreement with the CFM by starting from the Schwarzschild ($a=0$) result for our initial guess of the Kerr ($a\neq 0$) QNM solution using FindRoot in Mathematica in our AIM code (at least for $a\leq 1$).}
	
\par As presented in Tables \ref{tab:Blake51} and \ref{tab:Blake52} are the QNM frequencies for the scalar perturbations of the Kerr black hole with the two ``extreme'' (minimum and maximum) values of the angular momentum per unit mass, that is, $a = 0.00$ and $a = 0.80$. $m$ was set to $0$, while $l$ was given values of $0$, $1$ and $2$ and $n$ varied accordingly.

\par Included in Table \ref{tab:Blake51} are the numerically determined QNM frequencies published by Leaver in 1985 \cite{Leaver1985}. The percentages bracketed under each QNM frequency via the AIM are the percentage differences between the calculated value and the numerical value published by Leaver.  With the exceptions of the QNM frequencies for $l = 0$, $n = 0$ and $l = 2$, $n = 2$, the AIM values correspond to the CFM up to four decimal places and even those ``anomalies'' differ by less than $0.30\%$. Proving, at least in this case, the AIM is a precise semi-analytical technique.

\par In Table \ref{tab:Blake52}, all three values were calculated in this work, even though published values are available for the third order WKB(J), at least graphically, where numerical values using the CFM were taken from Ref. \cite{Berti:2005eb}. Since the WKB(J) is a generally accepted semi-analytical technique for QNM frequency calculations, the percentages below the AIM values are the differences to the sixth order WKB(J) values. Only in the case of $l = 0$, $n = 0$ does the AIM QNM frequency significantly differ from the sixth order WKB(J) value. Note that in an upcoming work, further values will be presented for values of $a = 0.20$, $a = 0.40$ and $a = 0.60$, with the same variations of $l$ and $n$ with $M=1$ \cite{Blake}.

%==================================================
\begin{table}[ht]
\caption{\textit{The QNM Frequencies for the Scalar Perturbations of the Kerr Black Hole, with $a = 0.00$, that is, the Schwarzschild limit ($M=1,~m=0$). Numerical data via the CFM taken from \cite{Berti:2005eb}, where the AIM was set to run at $15$ iterations.}}
\label{tab:Blake51}
\begin{ruledtabular}
\begin{tabular}{c|c|c|c|c|c|}
l & n & Numerical & Third Order WKB(J) & Sixth Order WKB(J) & AIM \\
\hline
0 & 0 & 0.1105 - 0.1049i & 0.1046 - 0.1001i & 0.1105 - 0.1008i & 0.1103 - 0.1046i \\
&   &   & (-5.34\% , 9.82\%) & ($<$0.01\% , -2.91\%) & (-0.18\% , -0.29\%) \\
\hline
1 & 0 & 0.2929 - 0.0977i & 0.2911 - 0.0989i & 0.2929 - 0.0978i & 0.2929 - 0.0977i \\
&   &   & (-0.61\% , 1.23\%) & ($<$0.01\% , 0.10\%) & ($<$0.01\% , $<$0.01\%) \\
& 1 & 0.2645 - 0.3063i & 0.2622 - 0.3074i & 0.2645 - 0.3065i & 0.2645 - 0.3063i \\
&   &   & (-0.87\% , 0.36\%) & ($<$0.01\% , 0.07\%) & ($<$0.01\% , $<$0.01\%) \\
\hline
2 & 0 & 0.4836 - 0.0968i & 0.4832 - 0.0968i & 0.4836 - 0.0968i & 0.4836 - 0.0968i \\
&   &   & (-0.08\% , $<$0.01\%) & ($<$0.01\% , $<$0.01\%) & ($<$0.01\% , $<$0.01\%) \\
& 1 & 0.4639 - 0.2956i & 0.4632 - 0.2958i & 0.4638 - 0.2956i & 0.4639 - 0.2956i \\
&   &   & (-0.15\% , 0.07\%) & (-0.02\% , $<$0.01\%) & ($<$0.01\% , $<$0.01\%) \\
& 2 & 0.4305 - 0.5086i & 0.4317 - 0.5034i & 0.4304 - 0.5087i & 0.4306 - 0.5086i \\
&   &   & (0.28\% , -1.02\%) & (-0.02\% , 0.02\%) & (0.02\% , $<$0.01\%)
\end{tabular}
\end{ruledtabular}
\end{table}
%==================================================
%==================================================
\begin{table}[ht]
\caption{\textit{The QNM Frequencies for the Scalar Perturbations of the Kerr Black Hole, with $a = 0.80$ ($M=1,~m=0$). Numerical data via the CFM taken from \cite{Berti:2005eb}, where the AIM was set to run at $15$ iterations.}}
\label{tab:Blake52}
\begin{ruledtabular}
\begin{tabular}{c|c|c|c|c|c|}
l & n & Numerical & Third Order WKB(J) & Sixth Order WKB(J) & AIM \\
\hline
0 & 0 & 0.1145 -0.0957i & 0.1005 - 0.1007i & 0.1211 - 0.0897i & 0.1141 - 0.0939i \\
&   &  &(-12.2\% , 5.22\%)  & (5.76\% ,- 6.23\%)    & (-0.35\% , -1.88\%) \\
\hline
1 & 0 &0.3067 -0.0901i & 0.3029 - 0.0891i & 0.3053 - 0.0893i & 0.3052 - 0.0892i \\
&   &  &(-1.24\% , -1.11\%) &  (-0.46\% , -0.89\%) & (-0.49\% , -1.00\%) \\
&1 & 0.2820 -.2783i & 0.2758 - 0.2779i & 0.2821 - 0.2755i & 0.2817 - 0.2756i \\
&   &  & (-2.20\%,-0.14\%) & (0.04\%, -1.01\%) & (-0.11\% , -0.97\%) \\
\hline
2 & 0 &0.5071 -0.0897i & 0.5035 - 0.0885i & 0.5041 - 0.0886i & 0.5041 - 0.0886i \\
&   &   & (-0.71\%, -1.34\%) & (-0.59\%, -1.23\%) & (-0.59\% , -1.23\%) \\
& 1 & 0.4906 -0.2722i & 0.4866 - 0.2693i & 0.4885 - 0.2690i & 0.4885 - 0.2689i \\
&   &   & (-0.82\%, -1.07\%) & (-0.43\%, -1.18\%) & (-0.43\% , -1.21\%) \\
& 2 & 0.4609 -0.4634i& 0.4585 - 0.4570i & 0.4607 - 0.4581i & 0.4606 - 0.4579i \\
& &   & (-0.52\%, -1.38\%)  & (-0.04\%, -1.14\%) & (-0.06\%, -1.19\%)
\end{tabular}
\end{ruledtabular}
\end{table}
%==================================================

%%%%%%%%%%%%%%%%%%%%%%%%%%%%%%%%%

\subsection{The Spin-Half Case}\label{sec:3-2}

\par For the spin-1/2 case we would like to know how the AIM can be used to derive an appropriate form of the Dirac equation in this spacetime background using the basis set up by four null vectors which are the basis of the Newman-Penrose formalism, for further details see Ref. \cite{Blake}. That is, in the Kerr background we adopt the following vectors as the null tetrad:
\begin{eqnarray}
l _ { j } & = & \frac { 1 } { \Delta } ( \Delta , - \rho ^ { 2 } , 0 , - a \Delta \sin ^ { 2 } { \theta } )\; ,\nonumber\\
n _ { j } & = & \frac { 1 } { 2 \rho ^ { 2 } } ( \Delta , \rho ^ { 2 } , 0 , - a \Delta \sin ^ { 2 } { \theta } )\; ,\nonumber\\
m _ { j } & = & \frac { 1 } { \sqrt { 2 } \bar { \rho } } ( i a \sin { \theta } , 0 , - \rho ^ { 2 } , - i ( r ^ { 2 } + a ^ { 2 } ) \sin { \theta } )\; ,\nonumber\\
l ^ { j } & = & \frac { 1 } { \Delta } ( r ^ { 2 } + a ^ { 2 } , \Delta , 0 , a )\; ,\nonumber\\
n ^ { j } & = & \frac { 1 } { \sqrt { 2 } \bar { \rho } } ( r ^ { 2 } + a ^ { 2 } , - \Delta , 0 , a )\; ,\nonumber\\
m ^ { j } & = & \frac { 1 } { \sqrt { 2 } \bar { \rho } } ( i a \sin { \theta } , 0 , 1 , - i \frac { 1 } { \sin { \theta } } )\; ,\nonumber
\end{eqnarray}
where $\bar { m } _ { j }$ and $\bar { m } ^ { j }$ are nothing but complex conjugates of $m _ { j }$ and $m ^ { j }$ respectively.

\par It is clear that the basis vectors basically become derivative operators when these are applied as tangent vectors to the function $e ^ { i ( \omega t + m \phi ) }$. Therefore we can write
\begin{equation}
\vec { l } = D = \mathcal { D } _ { 0 } \mbox { , } \vec { n } = D ^ { * } = - \frac { \Delta } { 2 \rho ^ { 2 } } \mathcal { D } ^ { \dagger } _ { 0 } \mbox { , } \vec { m } = \delta = \frac { 1 } { \sqrt { 2 } \bar { \rho } } \mathcal { L } ^ { \dagger } _ { 0 } \mbox { , } \vec { \bar { m } } = \delta ^ { * } = \frac { 1 } { \sqrt { 2 } \bar { \rho } ^ { * } } \mathcal { L } _ { 0 } \; ,
\end{equation}
where,
\begin{eqnarray}
\mathcal { D } _ { n } & = & \partial _ { r } + i \frac { K } { \Delta } + 2 n \frac { r - M } { \Delta } \; ,\nonumber\\
\mathcal { D } ^ { \dagger } _ { n } & = & \partial _ { r } - i \frac { K } { \Delta } + 2 n \frac { r - M } { \Delta } \; ,\nonumber\\
\mathcal { L } _ { n } & = & \partial _ { \theta } + Q + n \cot { \theta }\; ,\nonumber\\
\mathcal { L } ^ { \dagger } _ { n } & = &\partial _ { \theta } - Q + n \cot { \theta }\; ,\nonumber
\end{eqnarray}
and $K = ( r ^ { 2 } + a ^ { 2 } ) \omega + a m$ with $Q = a \omega \sin { \theta } + m \csc { \theta }$.

\par The spin coefficients can be written as a combination of basis vectors in the Newman-Penrose formalism which are now expressed in terms of the elements of different components of the Kerr metric. So by combining these different components of basis vectors in a suitable manner we get the spin coefficients as
\begin{equation}
\kappa = \sigma = \lambda = \nu = \varepsilon = 0 \; .
\end{equation}
\begin{eqnarray}
\tilde{ \rho } = - \frac { 1 } { \bar { \rho } ^ { * } } \mbox { , } \beta & = & \frac { \cot { \theta } } { 2 \sqrt { 2 } \bar { \rho } ^ { * } } \mbox { , } \pi = \frac { i a \sin { \theta } } { \sqrt { 2 } ( \bar { \rho } ^ { * } ) ^ { 2 } }\; ,\nonumber\\
\tau = - \frac { i a \sin { \theta } } { \sqrt { 2 } \rho ^ { 2 } } \mbox { , } \mu & = & - \frac { \Delta } { 2 \bar { \rho } ^ { * } \rho ^ { 2 } } \mbox { , } \gamma = \mu + \frac { r - M } { 2 \bar \rho ^ { 2 } }\; ,\\
\alpha & = & \pi - \beta ^ { * }\; .\nonumber
\end{eqnarray}
Using the above definitions, and by choosing: $f _ { 1 } = \bar { \rho } ^ { * } F _ { 1 }$, $g _ { 2 } = \bar { \rho } G _ { 2 }$, $f _ { 2 } = F _ { 2 }$ and $g _ { 1 } = G _ { 1 }$ (where $F_{1,2}$ and $G_{1,2}$ are a pair of spinors) the Dirac equation reduces to
\begin{eqnarray}
\mathcal { D } _ { 0 } f _ { 1 } + \frac { 1 } { \sqrt { 2 } } \mathcal { L } _ { \frac { 1 } { 2 } } f _ { 2 } = 0 \; ,\nonumber\\
\Delta \mathcal { D } ^ { \dagger } _ { \frac { 1 } { 2 } } f _ { 2 } - \sqrt { 2 } \mathcal { L } ^ { \dagger } _ { \frac { 1 } { 2 } } f _ { 1 } = 0 \; ,\label{eq:Blake221}\\
\mathcal { D } _ { 0 } g _ { 2 } - \frac { 1 } { \sqrt { 2 } } \mathcal { L } ^ { \dagger } _ { \frac { 1 } { 2 } } f _ { 1 } = 0 \; ,\nonumber\\
\Delta \mathcal { D } ^ { \dagger } _ { \frac { 1 } { 2 } } g _ { 1 } + \sqrt { 2 } \mathcal { L } _ { \frac { 1 } { 2 } } g _ { 2 } = 0 \; .\nonumber
\end{eqnarray}
We separate the Dirac equation into radial and angular parts by choosing,
\begin{eqnarray}
f _ { 1 } ( r , \theta ) = R _ { - \frac { 1 } { 2 } } ( r ) S _ { - \frac { 1 } { 2 } } ( \theta ) \mbox { , } f _ { 2 } ( r , \theta ) = R _ { \frac { 1 } { 2 } } ( r ) S _ { \frac { 1 } { 2 } } ( \theta )\; ,\nonumber\\
g _ { 1 } ( r , \theta ) = R _ { \frac { 1 } { 2 } } ( r ) S _ { - \frac { 1 } { 2 } } ( \theta ) \mbox { , } g _ { 2 } ( r , \theta ) = R _ { - \frac { 1 } { 2 } } ( r ) S _ { \frac { 1 } { 2 } } ( \theta )\; .\nonumber
\end{eqnarray}
Replacing these $f _ { j }$ and $g _ { j }$ $( j = 1 , 2 )$ and using $\lambda$ as the separation constant, we get,
\begin{eqnarray}
\mathcal { L } _ { \frac { 1 } { 2 } } S _ { \frac { 1 } { 2 } } = - \lambda S _ { - \frac { 1 } { 2 } }\; ,\label{eq:Blake222}\\
\mathcal { L } ^ { \dagger } _ { \frac { 1 } { 2 } } S _ { - \frac { 1 } { 2 } } = \lambda S _ { \frac { 1 } { 2 } }\; ,\nonumber
\end{eqnarray}
\begin{eqnarray}
\Delta ^ { \frac { 1 } { 2 } } \mathcal { D } _ { 0 } R _ { - \frac { 1 } { 2 } } = \lambda \Delta ^ { \frac { 1 } { 2 } } R _ { \frac { 1 } { 2 } }\; ,\label{eq:Blake223}\\
\Delta ^ { \frac { 1 } { 2 } } \mathcal { D } ^ { \dagger } _ { 0 } \Delta ^ { \frac { 1 } { 2 } } R _ { \frac { 1 } { 2 } } = \lambda R _ { - \frac { 1 } { 2 } }\; ,\nonumber
\end{eqnarray}
where $2 ^ { \frac { 1 } { 2 } } R _ { - \frac { 1 } { 2 } }$ is redefined as $R _ { - \frac { 1 } { 2 } }$.

\par Eqs. (\ref{eq:Blake222}) and (\ref{eq:Blake223}) are the angular and radial Dirac equation respectively, in a coupled form with the separation constant $\lambda$ \cite{Chandrasekhar:1985kt}. Decoupling Eq. (\ref{eq:Blake222}) gives the eigenvalue/angular equation for spin half particles as
\begin{equation}
\left [ \mathcal { L } _ { \frac { 1 } { 2 } } \mathcal { L } ^ { \dagger } _ { \frac { 1 } { 2 } } + \lambda ^ { 2 } \right ] S _ { - \frac { 1 } { 2 } } = 0 \; ,
\end{equation}
and $S _ { \frac { 1 } { 2 } }$ satisfies the `adjoint' equation (obtained by replacing $\theta$ by $\pi - \theta$).\footnote{Note that this angular equation is that given by Eq. (\ref{kerrang}) for $s=1/2$ and hence using the method in Appendix \ref{sec:A}, we could solve this numerically.} Decoupling Eq. (\ref{eq:Blake223}) then gives the radial equation for spin half particles as
\begin{equation}
\left [ \Delta \mathcal { D } ^ { \dagger } _ { \frac { 1 } { 2 } } \mathcal { D } _ { 0 } - \lambda ^ { 2 } \right ] R _ { - \frac { 1 } { 2 } } = 0 \; , \label{eq:Blake224}
\end{equation}
and $\Delta ^ { \frac { 1 } { 2 } } R _ { \frac { 1 } { 2 } }$ satisfies the complex-conjugate equation. Furthermore, unlike the case of a scalar particle, a spin half particle is not capable of extracting energy from a rotating black hole, that is, there is no Penrose Process (superradiance) equivalent scenario \cite{Blake}.

\par Returning now to the AIM, recall that it shall work better on a compact domain, where we define a new variable $y ^ { 2 } = 1 - \frac { r _ { + } } { r }$, which ranges from $0$ at the event horizon $(r= r_{ + } )$ to $1$ at spatial infinity. It is then necessary to incorporate the boundary conditions, which expressed in the new compact domain is
\begin{equation}
\psi ( y ) = \left ( 1 - \frac { r _ { - } } { r _ { + } } ( 1 - y ^ { 2 } ) \right ) ^ { - \frac { 1 } { 2 }- i \sigma _ { - } } ( y ^ { 2 } ) ^ { \frac { 1 } { 2 } - \sigma _ { + } } ( 1 - y ^ { 2 } ) ^ { - r _ { + } \omega } e ^ { i \omega \frac { r _ { + } } { 1 - y ^ { 2 } } } \chi ( y ) \; ,
\end{equation}
where  we have defined
\begin{equation}
\psi=\sqrt\Delta R_{1/2}+ R_{-1/2}~
\end{equation}
and $\psi$ satisfies the WKB(J)-like equation:
\begin{equation}
{d^2 \psi\over dy^2} + (\omega^2-V)\psi=0
\end{equation}
with potential
\begin{equation}
V=\lambda^2  {\Delta\over \bar K^2}+ \lambda {d\over dx} \Big( {\sqrt\Delta \over \bar K} \Big)
\end{equation}
and  $\bar K =K/\omega= ( r ^ { 2 } + a ^ { 2 } )  + {a } m/\omega$ (for more details see \cite{Blake}).

By making the change of coordinates and change of functions, our equation takes the form
\begin{equation}
\chi ( y ) = \lambda _ { 0 } ( y ) + s _ { 0 } ( y ) \; ,\label{eq:Blake59}
\end{equation}
where as in Sub-Sec.\ref{sec:5-2} we have
\begin{equation}
\lambda _ { 0 } = - 2 \frac { 1 } { g } \frac { \mathrm { d } g } { \mathrm { d } y } - \frac { 1 } { f } \frac { \mathrm { d } f } { \mathrm { d } y } \; ,
\end{equation}
and
\begin{equation}
s _ { 0 } = - \frac { 1 } { g } \frac { \mathrm { d } ^ { 2 } g } { \mathrm { d } y ^ { 2 } } - \frac { 1 } { f } \frac { \mathrm { d } f } { \mathrm { d } y } \times \frac { 1 } { g } \frac { \mathrm { d } g } { \mathrm { d } y } - \frac { 1 } { f ^ { 2 } } \left ( \omega ^ { 2 } - V | _ { r = r _ { + } ( 1 - y ^ { 2 } ) ^ { - 1 } } \right ) \; .
\end{equation}

\par As presented in Table \ref{tab:Blake53} and Table \ref{tab:Blake54} the QNM frequencies for the spin half perturbations of the Kerr black hole with the two ``extreme'' values of the angular momentum per unit mass, that is $a = 0.00$ and $a = 0.80$, $m$ was set to $0$, while $l$ was given values of $0$, $1$ and $2$ and $n$ varied accordingly.

\par Included in Table \ref{tab:Blake53} are the numerically determined QNM frequencies published by Jing {\it et al.} \cite{Jing:2005pk}. Even though the WKB method has been used to calculate the Schwarzschild limit QNM frequencies before \cite{Cho:2003qe}, the sixth order WKB values and AIM values are novel to this work and shall be explored more fully in Ref. \cite{Blake}. The percentages bracketed under each QNM frequency, is the percentage difference between the calculated value and the numerical value published by Jing {\it et al.} \cite{Jing:2005pk}.

\par As expected, since there are additional correction terms, the sixth order WKB QNM frequencies are closer to the numerical values than the third order WKB values. While the AIM does not prove as accurate in its calculation of the spin half QNM frequencies as it did with the scalar values (both for $15$ iterations), none of the differences between the AIM values and the numerical values exceed $0.30\%$, except for when $l = 2$, $n = 2$ (better accuracy can be achieved by increasing the number of iterations). 

\par Similarly in Table \ref{tab:Blake54} are the numerically determined QNM frequencies published by Jing {\it et al.} \cite{Jing:2005pk}. Both the third and sixth order WKB values along with the AIM values are novel to this work and shall also be explored more fully in Ref. \cite{Blake}. The percentages bracketed under each QNM frequency, are the percentage differences between the calculated value and the numerical value published by Jing {\it et al.}, at least for $l = 0$ and $l = 1$. For $l = 2$, the AIM values are compared to the sixth order WKB values. As already noted, since there are additional correction terms, the sixth order WKB QNM frequencies are closer to the numerical values than the third order WKB values. Again the AIM does not appear to be as precise in calculating the QNM frequencies for spin half perturbations of the Kerr black hole as it was for the scalar perturbations (at least for $15$ iterations). As we mentioned, additional tables and plots of these Kerr processes shall constitute a future work \cite{Blake}.

%==================================================
\begin{table}[ht]
\caption{\textit{The QNM Frequencies for the Spin Half Perturbations of the Kerr black hole, with $a = 0.00$, that is, the Schwarzschild limit ($M=1,~m=0$). Numerical data via the CFM taken from \cite{Jing:2005pk}, where the AIM was set to run at $15$ iterations.}}
\label{tab:Blake53}
\begin{ruledtabular}
\begin{tabular}{|c|c|c|c|c|c|}
l & n & Numerical & Third Order WKB & Sixth Order WKB & AIM \\
\hline
0 & 0 & 0.1830 - 0.0970i & 0.1765 - 0.1001i & 0.1827 - 0.0949i & 0.1830 - 0.0969i \\
&   &   & (-3.55\% , 3.20\%) & (-0.16\% , -2.16\%) & ($<$0.01\% , -0.10\%) \\
\hline
1 & 0 & 0.3800 - 0.0964i & 0.3786 - 0.0965i & 0.3801 - 0.0964i & 0.3800 - 0.0964i \\
&   &   & (-0.37\% , 0.10\%) & (0.03\% , $<$0.01\%) & ($<$0.01\% , $<$0.01\%) \\
& 1 & 0.3558 - 0.2975i & 0.3536 - 0.2987i & 0.3559 - 0.2973i & 0.3568 - 0.2976i \\
&   &   & (-0.62\% , 0.40\%) & (0.03\% , -0.07\%) & (0.28\% , 0.03\%) \\
\hline
2 & 0 & 0.5741 - 0.0963i & 0.5737 - 0.0963i & 0.5741 - 0.0963i & 0.5741 - 0.0963i \\
&   &   & (-0.07\% , $<$0.01\%) & ($<$0.01\% , $<$0.01\%) & ($<$0.01\% , $<$0.01\%) \\
& 1 & 0.5570 - 0.2927i & 0.5562 - 0.2930i & 0.5570 - 0.2927i & 0.5573 - 0.2928i \\
&   &   & (-0.14\% , 0.10\%) & ($<$0.01\% , $<$0.01\%) & (0.05\% , 0.03\%) \\
& 2 & 0.5266 - 0.4997i & 0.5273 - 0.4972i & 0.5265 - 0.4997i & 0.5189 - 0.5213i \\
&   &   & (0.13\% , -0.50\%) & (-0.02\% , $<$0.01\%) & (-1.46\% , 4.32\%)
\end{tabular}
\end{ruledtabular}
\end{table}
%==================================================
%==================================================
\begin{table}[ht]
\caption{\textit{The QNM Frequencies for the Spin Half Perturbations of the Kerr black hole, with $a = 0.80$ ($M=1,~m=0$). Numerical data via the CFM taken from \cite{Jing:2005pk}, where the AIM was set to run at $15$ iterations.}}
\label{tab:Blake54}
\begin{ruledtabular}
\begin{tabular}{|c|c|c|c|c|c|}
l & n & Numerical & Third Order WKB(J) & Sixth Order WKB(J) & AIM \\
\hline
0 & 0 & 0.1932 - 0.0891i & 0.1883 - 0.0896i & 0.1914 - 0.0865i & 0.1920 - 0.0872i \\
&   &   & (-2.54\% , 0.56\%) & (-0.93\% , -2.92\%) & (-0.62\% , -2.13\%) \\
\hline
1 & 0 & 0.3993 - 0.0893i & 0.3956 - 0.0881i & 0.3967 - 0.0880i & 0.3965 - 0.0880i \\
&   &   & (-0.93\% , -1.34\%) & (-0.65\% , -1.46\%) & (-0.70\% , -1.46\%) \\
& 1 & 0.3789 - 0.2728i & 0.3751 - 0.2701i & 0.3777 - 0.2687i & 0.3764 - 0.2517i  \\
&   &   & (-1.00\% , -0.99\%) & (-0.32\% , -1.50\%) & (-0.66\%, -7.73\%)  \\
\hline
2 & 0 &   & 0.5984 - 0.0881i & 0.5987 - 0.0881i & 0.5987 - 0.0882i \\
&   &   &   &   & ($<$0.01\% , 0.11\%) \\
& 1 &   & 0.5844 - 0.2669i & 0.5855 - 0.2667i & 0.5846 - 0.2644i \\
&   &   &   &   & (-0.15\% , -0.86\%) \\
& 2 &   & 0.5600 - 0.4512i & 0.5609 - 0.4517i & 0.6023 - 0.4260i  \\
&   &   &   &   & (7.38\%, -5.70\%)
\end{tabular}
\end{ruledtabular}
\end{table}
%==================================================

%%%%%%%%%%%%%%%%%%%%%%%%%%%%%%%%%

\subsection{The Spin-Two Case}
\label{sec:5-3}

\par As we have mentioned earlier, the radial equation for nonzero spin $s$ is in general complex. In fact, it does not even reduce to the Regge-Wheeler and Zerilli equations when the rotation parameter $a\rightarrow 0$. Detweiler \cite{Detweiler1977} has found a way to overcome this problem, where he defined a new function
\begin{equation}
X=\Delta^{s/2}\left(r^{2}+a^{2}\right)^{1/2}\left[\alpha(r)R+\beta(r)\Delta^{s+1}\frac{dR}{dr}\right] \; .
\end{equation}
If the functions $\alpha(r)$ and $\beta(r)$ are required to satisfy
\begin{equation}
\alpha^{2}-\alpha'\beta\Delta^{s+1}+\alpha\beta'\Delta^{s+1}-\beta^{2}\Delta^{2s+1}K={\rm constant} \; ,
\end{equation}
then it can be shown that the radial equation in Eq. (\ref{kerrrad}) becomes
\begin{equation}
\frac{d^{2}X}{dx^{2}}-VX=0\; ,\label{xeqn}
\end{equation}
where
\begin{eqnarray}
V&=&\frac{\Delta U}{\left(r^{2}+a^{2}\right)^{2}}+G^{2}+\frac{dG}{dx}\; ,\\
G&=&\frac{s(2r-1)}{2\left(r^{2}+a^{2}\right)}+\frac{r\Delta}{\left(r^{2}+a^{2}\right)^{2}}\; ,\\
U&=&K+\frac{2\alpha'+\left(\beta'\Delta^{s+1}\right)'}{\beta\Delta^{s}}\; ,\\
x&=&r+\frac{r_{+}}{r_{+}-r_{-}}{\rm ln}\left(r-r_{+}\right)-\frac{r_{-}}{r_{+}-r_{-}}{\rm ln}\left(r-r_{-}\right)\; .
\end{eqnarray}
As Detweiler has indicated, it is possible to choose the functions $\alpha(r)$ and $\beta(r)$ so that the resulting effective potential $V(r)$ is real and has the form
\begin{eqnarray}
V&=&\frac{\rho^{2}\Delta}{\left(r^{2}+a^{2}\right)^{2}}\left\{\frac{f(f+2)}{g+b\Delta}-\frac{b\Delta}{\rho^{4}}
+\frac{\left[\kappa\rho\Delta-\left(g'\Delta-g\Delta'\right)\right]\left[\kappa\rho g-b\left(g'\Delta-g\Delta'\right)\right]} {\rho^{2}\left(g+b\Delta\right)\left(g-b\Delta\right)^{2}}\right\}\nonumber\\
&&\ \ \ +\left[\frac{ram\Delta}{\omega\rho\left(r^{2}+a^{2}\right)^{2}}\right]^{2}
-\frac{\Delta}{r^{2}+a^{2}}\frac{d}{dr}\left[\frac{ram\Delta}{\omega\rho\left(r^{2}+a^{2}\right)^{2}}\right]-\frac{Y^{2}}{\left(r^{2}+a^{2}\right)^{2}}\; , \label{realpot}
\end{eqnarray}
where
\begin{eqnarray}
g&=&f\rho^{2}+3\rho(r^{2}+a^{2})-3r^{2}\Delta\; ,\\
\rho&=&r^{2}+a^{2}-\frac{am}{\omega}\; ,\\
\kappa&=&\pm\left\{9-2f\left[\left(a^{2}-\frac{am}{\omega}\right)(5f+6)-12a^{2}\right]+2bf(f+2)\right\}^{1/2}\; ,\\
b&=&\pm3\left(a^{2}-\frac{am}{\omega}\right)\; ,\label{beqn}\\
Y&=&am-(r^{2}+a^{2})\omega\; ,\\
f&=&A+a^{2}\omega^{2}-2am\omega\; .
\end{eqnarray}
When the Kerr rotation parameter $a$ approaches zero, the potential $V$ in Eq. (\ref{realpot}) coincides with the Regge-Wheeler potential for negative $\kappa$, and coincides with the Zerilli potential for positive $\kappa$. Here we choose $\kappa$ to be negative, where the choice of the sign in Eq. (\ref{beqn}) is determined by the sign of $m$ \cite{Detweiler1977,Seidel:1989bp}.

\par The QNM boundary conditions for $X$ are
\begin{equation}
X\rightarrow\left\{
\begin{array}{cl} e^{-ikx} & ; \ x\rightarrow-\infty \\ e^{i\omega x} & ; \ x\rightarrow\infty
\end{array}\right.\; ,
\end{equation}
where
\begin{equation}
k=\omega-\frac{am}{r_{+}}\; .
\end{equation}
Hence, we write
\begin{equation}
X=e^{i\omega r}r^{\frac{i\omega}{r_{+}-r_{-}}}\left[\frac{(r-r_{+})^{r_{+}}}{(r-r_{-})^{r_{-}}}\right]^{-\frac{ik}{r_{+}-r_{-}}}\chi_{G}\; .
\end{equation}
Substituting this into Eq. (\ref{xeqn}), we have
\begin{eqnarray}
&&\chi_{G,rr}+\left[2\Gamma_{G}+\frac{r^{2}-a^{2}}{\Delta\left(r^{2}+a^{2}\right)}\right]\chi_{G,r}
+\left[\Gamma_{G}^{2}+\Gamma_{G,r}+\frac{\Gamma_{G}\left(r^{2}-a^{2}\right)}{\Delta\left(r^{2}+a^{2}\right)} -\left(\frac{r^{2}+a^{2}}{\Delta}\right)^{2}V\right]\chi_{G}=0\; ,\label{chieqn}
\end{eqnarray}
where
\begin{equation}
\Gamma_{G}=i\omega+\frac{i\omega}{r(r_{+}-r_{-})}-\frac{ikr}{\Delta}\; .
\end{equation}
As we did earlier, we define the variable $\xi=1-r_{+}/r$ which has a compact domain $0<\xi<1$. Eq. (\ref{chieqn}) can then be written in the AIM form
\begin{equation}
\chi_{G,\xi\xi}=\lambda_{G}(\xi)\chi_{G,\xi}+s_{G}(\xi)\chi_{G}\; ,
\end{equation}
where
\begin{eqnarray}
\lambda_{G}&=&\frac{2}{1-\xi}-\frac{r_{+}}{(1-\xi)^{2}}\left[2\Gamma_{G} +\frac{1}{\Delta}\frac{r_{+}^{2}-a^{2}(1-\xi)^{2}}{r_{+}^{2}+a^{2}(1-\xi)^{2}}\right]\; ,\\
s_{G}&=&\frac{r_{+}^{2}}{(1-\xi)^{4}}\left\{\left[\frac{r_{+}^{2}+a^{2}(1-\xi)^{2}}{\Delta(1-\xi)^{2}}\right]^{2}V
-\frac{\Gamma_{G}}{\Delta}\frac{r_{+}^{2}-a^{2}(1-\xi)^{2}}{r_{+}^{2}+a^{2}(1-\xi)^{2}}-\Gamma_{G}^{2}\right\} -\frac{r_{+}}{(1-\xi)^{2}}\Gamma_{G,\xi}\; .
\end{eqnarray}

\par The results for the gravitational (spin-two) case are presented in Tables~\ref{tab:Blake55S} and \ref{tab:Blake56S}. In general the error in the separation constant is smaller than that of the quasinormal frequencies. As for the quasinormal frequencies the error in the Kerr case is larger than that of either the Schwarzschild or the Reissner-Nordstr\"om cases, where this is due to our consideration of the angular and the radial equations simultaneously. The number of iterations that can be performed in the code is relatively small, much like the number of continued fractions in the CFM is typically smaller due to the coupling between radial and angular equations.

%==================================================
\begin{table}[t]
\caption{\textit{Spin-2 angular separation constants and Kerr gravitational quasinormal frequencies for the fundamental mode corresponding to $l=2$ and $m=0$ compared with the CFM \cite{Leaver1985} ($M=1/2$).}}
\label{tab:Blake55S}
\begin{ruledtabular}
\begin{tabular}{|c|c|c|c|c|}
$a$ & $A_{Leaver}$ & $A_{AIM}$ & $\omega_{Leaver}$ & $\omega_{AIM}$ \\
\hline
0 & (4.0000, 0.0000) & (4.0000, 0.0000) & (0.7473, -0.1779)  & (0.7413, -0.1780) \\
&   & ($<$0.01\%)($<$0.01\%)  &   & (-0.80\%)(-0.06\%) \\
\hline
0.1 & (3.9972, 0.0014) & (3.9973, 0.0014) & (0.7502, -0.1774) & (0.7444, -0.1775) \\
&   & ($<$0.01\%)($<$0.01\%)  &   & (-0.77\% , -0.06\%) \\
\hline
0.2 & (3.9886, 0.0056) & (3.9887, 0.0056) & (0.7594, -0.1757) & (0.7540, -0.1763) \\
&   & ($<$0.01\%)($<$0.01\%)  &   & (-0.71\%)(-0.03\%) \\
\hline
0.3 & (3.9730, 0.0126) & (3.9733, 0.0126) & (0.7761, -0.1720) & (0.7715, -0.1722) \\
&   & ($<$0.01\%)($<$0.01\%)  &   & (-0.59\%)(-0.12\%) \\
\hline
0.4 & (3.9480, 0.0223) & (3.9482, 0.0222) & (0.8038, -0.1643) & (0.8025, -0.1639) \\
&   &  ($<$0.01\%)(-0.45\%) &   & (-0.16\%)(0.24\%) \\
\hline
0.45 & (3.9304, 0.0276) & (3.9303, 0.0280) & (0.8240, -0.1570)  & (0.8250, -0.1591) \\
 &   & ($<$0.01\%)(1.45\%)  &   & (0.12\%)(-1.34\%)
\end{tabular}
\end{ruledtabular}
\end{table}
%==================================================
%==================================================
\begin{table}[t]
\caption{\textit{Spin-2 angular separation constants and Kerr gravitational quasinormal frequencies for the fundamental mode corresponding to $l=2$ and $m=1$ compared with the CFM \cite{Leaver1985} ($M=1/2$).}}
\label{tab:Blake56S}
\begin{ruledtabular}
\begin{tabular}{|c|c|c|c|c|}
$a$ & $A_{Leaver}$ & $A_{AIM}$ & $\omega_{Leaver}$ & $\omega_{AIM}$ \\
\hline
0 & (4.0000, 0.0000) & (4.0000, 0.0000) & (0.7473, -0.1779)  & (0.7413, -0.1780) \\
&   & ($<$0.01\%)($<$0.01\%)  &   & (-0.80\%)(-0.06\%) \\
\hline
0.1 & (3.8932, 0.0252) & (3.8937, 0.0250) & (0.7765, -0.1770) & (0.7726, -0.1755) \\
&   & (-0.01\%)(-0.89\%)  &   & (-0.51\% , 0.86\%) \\
\hline
0.2 & (3.7676, 0.0532) & (3.7681, 0.0526) & (0.8160, -0.1745) & (0.8143, -0.1726) \\
&   & (0.01\%)(-1.12\%)  &   & (-0.02\%)(1.09\%) \\
\hline
0.3 & (3.6125, 0.0835) & (3.6123, 0.0826) & (0.8719, -0.1693) & (0.8722, -0.1674) \\
&   & ($<$0.01\%)(-0.99\%)  &   & (0.03\%)(1.00\%) \\
\hline
0.4 & (3.4023, 0.1122) & (3.4011, 0.1110) & (0.9605, -0.1559) & (0.9620, -0.1543) \\
&   &  (-0.03\%)(-1.00\%) &   & (0.16\%)(1.05\%) \\
\hline
0.45 & (3.2535, 0.1195) & (3.2491, 0.1173) & (1.0326, -0.1396)  & (1.0376, -0.1369) \\
 &   & (-0.03\%)(-1.82\%)  &   & (0.48\%)(1.97\%)
\end{tabular}
\end{ruledtabular}
\end{table}
%==================================================

%%%%%%%%%%%%%%%%%%%%%%%%%%%%%%%%%
\section{Doubly Rotating Kerr (A)dS Black Holes}\label{sec:6}\cleqn

\par Rotating black holes in higher dimensions were first discussed in the seminal paper by Myers and Perry \cite{Myers:1986un}.  One of the unexpected results to come from this work was that some families of solutions were shown to have event horizons for arbitrarily large values of their rotation parameters. The stability of such black holes is certainly in question \cite{Emparan:2003sy,Konoplya:2011qq}, with numerical evidence recently provided by Shibata and Yoshino \cite{Shibata:2009ad}. 

\par Another new feature of the Myers-Perry (MP) solutions is that they in general have $\lfloor\frac{D-1}{2}\rfloor$ spin parameters, making them more complex than the four dimensional Kerr solution. The first asymptotically non-flat five-dimensional MP metric was given in Ref. \cite{Hawking:1998kw}. Subsequent generalizations to arbitrary dimensions was done in Ref. \cite{Gibbons:2004js}, and finally the most general Kerr-(A)dS-NUT metric was found by Chen, L\"u and Pope \cite{Chen:2006xh}.
 
\par In this section we review how the AIM can be used to solve the $D\geq 6$ two-rotation scalar perturbation equations (for more details on the metric and resulting separation see Ref. \cite{Cho:2011yp}). The scalar field master equations are found to be \cite{Cho:2011yp}:
\begin{eqnarray}
  0&=&\frac{1}{r^{D-6}}\frac{d}{dr}\left(r^{D-6}\Delta_{r} \frac{d R_r}{dr}\right)+\left( \frac{(r^2+a_1^2)^2(r^2+a_2^2)^2}{\Delta_r}\tilde{\omega}_r^2-\frac{a_1^2 a_2^2 j(j+D-7)}{r^2}-b_1 r^2 -b_2\right) R_r\;, \label{eqn:Rr}\\
  0&=&\left(\frac{a_i}{y_i}\right)^{ D-6} \frac{d}{dy_i}\left[\left(\frac{y_i}{a_i}\right)^{D-6}\Delta_{y_i} \frac{d R_{\theta_i}}{dy_i}\right]-\left\{ \frac{(a_1^2-y_i^2)^2(a_2^2-y_i^2)^2}{\Delta_{y_i}}\tilde{\omega}_{y_i}^2+\frac{a_1^2 a_2^2 j(j+D-7)}{y_i^2}+b_1 y_i^2 -b_2\right\} R_{\theta_i} \; , \nonumber \\
  && \label{eqn:Ryi}
\end{eqnarray}
where 
\begin{eqnarray}
\Delta_{r}&=&(1+g^{2}r^{2})(r^{2}+a_{1}^{2})(r^{2}+a_{2}^{2})-2Mr^{7-D},\\
\Delta_{y_{i}}&=&(1-g^{2}y_{i}^{2})(a_{1}^{2}-y_{i}^{2})(a_{2}^{2}-y_{i}^{2})\;,
\end{eqnarray}
the radial and angular frequencies are defined by:
\begin{eqnarray}
\label{suprad}
  \tilde{\omega}_r&=&\omega-(1+g^2 r^2) \left(\frac{m_1 a_1}{r^2+a_1^2}+\frac{m_2 a_2}{r^2+a_2^2}\right),\\
  \tilde{\omega}_{y_i}&=&\omega-(1-g^2 y_i^2) \left(\frac{m_1 a_1}{a_1^2-y_i^2}+\frac{m_2 a_2}{a_2^2-y_i^2}\right),
\end{eqnarray}
and $i=1,2$. In the above $g$ is the curvature of the spacetime satisfying $R_{\mu\nu}=-3g^2 g_{\mu\nu}$ (e.g., see Ref. \cite{Chen:2006xh}), and $a_1,a_2$ are the two rotation parameters and for later reference we define $\epsilon=a_2/a_1$. 

\par Doubly rotating black holes are more complicated than simply rotating black holes (cf. Ref. \cite{Kodama:2009rq}), because two rotation planes lead to two coupled spheroids which are also needed for the solution of the radial equation. 

%%%%%%%%%%%%%%%%%%%%%%%%%%%%%%%%%%%%%%%%%%%%

\subsection{Radial Quasi-Normal Modes}\label{sec:6-1}

\par For simplicity we will consider the flat case, setting $g=0$, which leads to easier QNM boundary conditions (cf. Schwarzschild to Schwarzschild-dS). These satisfy the boundary condition that there are only waves ingoing at the black hole horizon and outgoing waves at asymptotic infinity.

\par As we have shown with the previous examples, it is easier to work on a compact domain and define the variable $x=1/r$, so that infinity is mapped to zero and the outer horizon stays at $x_h=1/r_h=1$. The domain of $x$ will therefore be $[0,1]$. Thus the QNM boundary condition is translated into the statement that the waves move leftward at $x=0$ and rightward at $x=1$.  We again choose the AIM point in the middle of the domain, that is, at $x=1/2$.

\par In terms of $x$ the radial equation (\ref{eqn:Rr}) becomes:
\begin{equation}\label{eqn:radialx}
  0=-x^{D-4}\frac{d}{dx}\left(-x^{8-D}\Delta_{x} \frac{d R}{dx}\right)+\left( \frac{(x^{-2}+a_1^2)^2(x^{-2}+a_2^2)^2}{\Delta_x}\tilde{\omega}_x^2-a_1^2 a_2^2 j(j+D-7)x^2-\frac{b_1}{x^2} -b_2\right) R\;, 
\end{equation}
where $\Delta_x(x)\equiv\Delta_r(r=1/x)$ and $\omega_x(x)\equiv\omega_r(r=1/x)$.

\par After performing some asymptotic analysis, we find that for the solutions to satisfy the QNM boundary conditions we must have:
\begin{equation}\label{eqn:y}
  R \sim (1-x)^{i \tilde{\omega}_h\alpha_h} x^{(D-2)/2}e^{i \omega_x/x} y(x)\;, 
\end{equation}
where 
\begin{eqnarray}
  \tilde{\omega}_h&\equiv&\omega_x(x=1)\;,\\
  \alpha_h&\equiv&\frac{(1+a_1^2)(1+a_2^2)}{\Delta_x'(x=1)}\;.
\end{eqnarray}
We then substitute this ansatz into Eq. (\ref{eqn:radialx}) and rewrite into the AIM form:
\begin{equation}
y''=\lambda_0 y' +s_0 y\;.
\end{equation}
This final step above can be performed in Mathematica, where the resulting expressions for $\lambda_0$ and $s_0$ are fed into the AIM routine. The method we use to find the QNMs proceeds in a fashion similar to that used in Ref. \cite{Berti:2005gp,Berti:2005ys} (see also Sec \ref{sec:5-3}) except we use the AIM instead of the CFM.

\par First we set the number of AIM iterations in both the eigenvalue and QNM calculations to sixteen. We start with the Schwarzschild values $(b_1,b_2,\omega)$, that is, at the point $(a_1,a_2)\sim 0$  and then increment $a_1$ and $a_2$ by some small value. We take the initial eigenvalues $(b_1,b_2)$, insert them into the radial equation (\ref{eqn:y}) then use the AIM to find the new QNM that is closest to $\omega$ using the Mathematica routine {\tt  FindRoot}. 

%==================================================
\begin{figure}[t]
  \centering
  \includegraphics[scale=1.25]{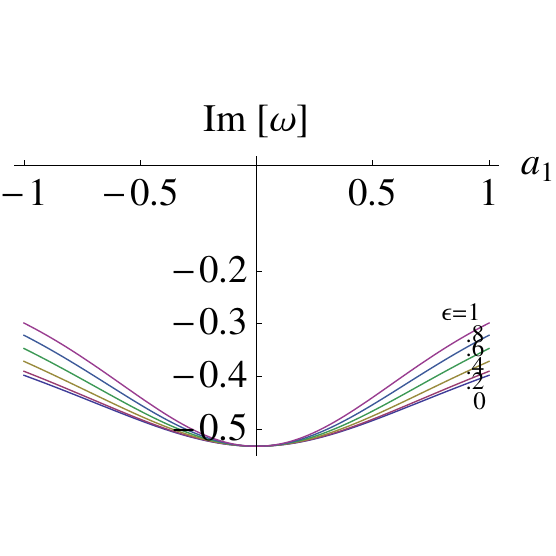}
  \hspace{1cm}
  \includegraphics[scale=1.25]{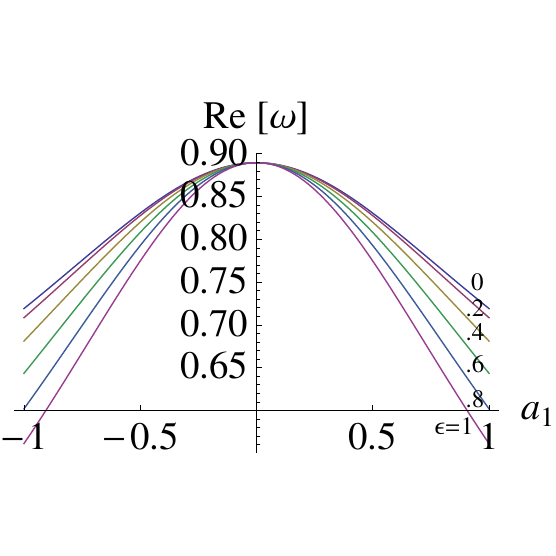}
\caption{An example of the $D=6$ fundamental $(j,m_1,m_2,n_1,n_2)=(0,0,0,0,0)$ QNM. On the left is a plot of the imaginary part and on the right plot of the real part.
\label{fig:D6000plots}
}
\end{figure}
%==================================================

\par Taking this new value of omega, $\omega'$, we insert it into the two angular equations (at the same value of $a_1$ and $a_2$) then solve using the AIM, searching closest to the previous $b_1$ and $b_2$ values. Thereby obtaining the new eigenvalues $b_1'$, $b_2'$. We then repeat this process with the new $(\omega',b_1',b_2')$ as the starting point until the results converge and we have achieved four decimal places of accuracy. When this occurs we increment $a_1$ and $a_2$ again and repeat the process. In this way, we are able to find the QNMs and eigenvalues along lines passing approximately through the origin (that is, starting from the near Scwharzschild values) in the $(a_1,a_2)$ parameter space. 

\par As an example, we have plotted various values of $\epsilon=a_2/a_1$ ($=0,0.2,0.4,0.6,0.8,1$) against $a_1$ and used an interpolating function to interpolate between these values as shown in Fig. \ref{fig:D6000plots} (for further details see Ref. \cite{Cho:2011yp}).

%%%%%%%%%%%%%%%%%%%%%%%%%%%%%%%%%
%
% Section 7: Summary and Outlook
%

\section{Summary and Outlook}\label{sec:7}\cleqn
\par In this review we have shown that the AIM can be used to calculate the radial QNMs of a variety of black hole spacetimes. In particular, we have used it to calculate perturbations of Schwarzschild (in asymptotically flat, de Sitter and anti-DeSitter), RN and Kerr (for spin $0, 1$ and $2$ perturbations) black holes in four dimensions. We argued that the method will be of use in studies of extra dimensional black holes and gave an explicit example of this in the case of the doubly rotating Myers Perry black hole.

\par We have hopefully demonstrated how the AIM can also be applied to radial QNMs and not just to spheroidal eigenvalue problems \cite{Barakat:2006ki,Cho:2009wf}. Given the fact that the AIM can be used in both the radial and angular wave equations \cite{Cho:2009wf} we expect no problems in obtaining QNMs for Kerr-dS black holes in four and higher dimensions. Note that this was only recently accomplished via the CFM in Ref. \cite{Yoshida:2010zzb} using Heun's equation \cite{Suzuki:1998vy} to reduce the problem to a 3-term recurrence relation. In higher dimensions a similar method was used for simply rotating Kerr-AdS black holes \cite{Kodama:2009rq}.

\par It remains to be seen if the AIM can be tailored to handle asymptotic QNMs (see Ref. \cite{Nollert:1993zz} for an adapted version of the CFM). However, given the close relationship between the AIM and the exact WKB approach \cite{Matamala:2007}, it might be possible to adapt the AIM to find asymptotic QNMs \cite{Motl:2003cd,Andersson:2003fh,Das:2004db,Ghosh:2005aq} numerically or even semi-analytically.

\par The AIM might be of some topical use, for example, in the angular spheroids/QNMs needed in the phenomenology of Hawking radiation from spinning higher-dimensional black holes, for a recent review see Ref. \cite{Frost:2009cf}. We recently used a combination of all the techniques discussed in this work to evaluate the angular eigenvalues, ${}_0A_{kjm}$, for real $c=a\omega$, which are needed for the tensor graviton emission rates on a {\it simply} rotating Kerr-de Sitter black hole background in $(n+4)$-dimensions \cite{Doukas:2009cx} (also see Ref. \cite{Kanti:2009sn}) and it might also be interesting to find QNMs of doubly rotating Kerr-(A)dS black holes (for asymptotically flat Kerr see Ref. \cite{Cho:2011yp}). Finally, attempting to solve the QNMs for all spins on the Schwarzschild-AdS background via the AIM also seems an interesting problem.

\par As such we hope to have provided the reader with enough technical details, and to have addressed some of the possible questions to allow them to pursue the study of QNMs with the AIM.\footnote{Source code and other information for some of the cases presented here can be found on the AIM link at \url{http://www-het.phys.sci.osaka-u.ac.jp/~naylor/}.}

%%%%%%%%%%%%%%%%%%%%%%%%%%%%%%%%%
%
% Acknowledgements
%

\acknowledgments

\par  HTC was supported in part by the National Science Council of the Republic of China under the Grant  NSC 99-2112-M-032-003-MY3, and the National Science Centre for Theoretical Sciences. The work of JD was supported by the Japan Society for the Promotion of Science (JSPS), under fellowship no. P09749. WN would like to thank the Particle Physics Theory Group, Osaka University for computing resources.

%%%%%%%%%%%%%%%%%%%%%%%%%%%%%%%%%%%%%%%%%%%%

\appendix

\section{Angular Eigenvalues for Spin-Weighted Spheroidal Harmonics}
\label{sec:A}

%==================================================
\begin{table}[t]
\caption{\sl Selected spin two eigenvalues, ${}_2A_{lm}$, obtained from the AIM for a Kerr black hole with varying values of $c=a\omega$ and $m=1$. The same number of iterations in the AIM, $n_A$ and the number of recursions in the CFM, $n_C$ for $c=0.1, 0.8$ at a working precision of $15$ digit precision, where results are presented to 10 s.f.}
\label{tab:spinTwo}
\begin{ruledtabular}
\begin{tabular}{|c|c|c|c|c|}
$l$ &  $c=0.1$ (n${}_A$=n${}_C$=15,  ) & $c=0.8$ (n${}_A$=35, n${}_C$=70) & $c=-10 i$ (n${}_A$=80, n${}_C$=130 CFM)  & $c=10 $  (n${}_A$=75, n${}_C$=145)\\
\hline
2 &   -0.1391483511& -1.462479552 & (12.44128209, 0.8956143162) & -101.8949078\\
3 &    5.929826236 & 5.247141863 & (32.31138608, 1.302608040) &-63.74900642\\
4 &    13.95640426 & 13.45636668  & (51.27922784, 1.946041848) &-30.35607486\\
5 &    23.96944247& 23.54163307 & (69.25750923,  3.012877154) &-4.557015739\\
6 &    35.97681567 & 35.58524928 & (85.86796852, 4.990079008) &-2.555206382\\
7 &    9.98139515 & 49.61077286 & (99.20081385, 6.801617108)  &12.32203552 \\
\end{tabular}
\end{ruledtabular}
\end{table}
%==================================================

\par As mentioned in Sec. \ref{sec:5-3}, aside from radial QNMs the AIM can also be applied to various kinds of  spin-weighted spheroidal harmonics, ${}_sS_{lm}(\theta)$, e.g. see Ref. \cite{Berti:2005ys}. Therefore, in this appendix, we briefly compare the AIM with the CFM for the four-dimensional spin-weighted spheroids.

\par With the regular boundary conditions, the angular wave function can be written as \cite{Leaver1985}
\begin{equation}
S=e^{a\omega u}(1+u)^{\frac{1}{2}|m-s|}(1-u)^{\frac{1}{2}|m+s|}\chi_{A}(u)\; .
\end{equation}
Putting this back into Eq. (\ref{kerrang}) and rewriting the equation in AIM form, we have
\begin{equation}
\chi_{A,uu}=\lambda_{A}(u)\chi_{A,u}+s_{A}(u)\chi_{A}\; ,
\end{equation}
where
\begin{eqnarray}
\lambda_{A}(u)&=&\frac{2u}{1-u^{2}}-2N\; ,\\
s_{A}(u)&=&\frac{1}{1-u^{2}}\left[\frac{(m+su)^{2}}{1-u^{2}}+2a\omega su+2uN-\left(a^{2}\omega^{2}u^{2}+s+{}_sA_{lm}\right)\right]-N^{2}-N_{,u}\; ,\\
N&=&a\omega+\frac{|m-s|}{2(1+u)}-\frac{|m+s|}{2(1-u)}\; .
\end{eqnarray}
These are the relevant equations for calculating the eigenvalues of the spheroidal harmonics in the four-dimensional case. It was noticed in Ref. \cite{Barakat:2006ki} that the AIM converges fastest at the maximum of the potential, when the AIM is written in WKB form. In four dimensions this occurs at $x=0$ and is true for general spin-$s$ as we have verified.  Note that for higher dimensional generalizations it is not easy to explicitly find a maximum \cite{Cho:2009wf}. 

\par It may be worth mentioning that for the case where $c=0$ an exact analytic solution of the above equations leads to \cite{Berti:2005ys}:
\begin{equation}
{}_sA_{lm}= l(l+1)-s(s+1)\; .
\label{spherhar}
\end{equation}
In the exact limit $c=0$ the AIM recovers the result for spherical harmonics, Eq. (\ref{spherhar}) above, while for the CFM taking $c=0$  leads to singularities \cite{Berti:2005ys}; however for $c\ll 1$ we find agreement with the CFM and Eq. (\ref{spherhar}).

\par For the purposes of consistency we have calculated (see Table \ref{tab:spinTwo}) the ${}_2A_{l1}$  eigenvalues for the lowest $l=2,\dots,7$ modes to 10 significant figures and have also compared this with the results of the CFM. In both the AIM and CFM larger $l$ modes require more iterations/recursions to achieve convergence in a given $l$ eigenvalue to the required precision. Care should be taken when comparing the number of iterations in the AIM with that of the number of recursions in the CFM, because one iteration of the AIM is not equivalent to one iteration of the CFM. In fact although we typically need to iterate the improved AIM on average a lesser number of times, the CFM is typically faster  for smaller values of c. However, for larger values of $c$ both methods can be faster or slower.

\par The results of the first few $l$  eigenvalues for different values of $c=a\omega$, with $m=1$, are presented in Table \ref{tab:spinTwo}. As far as we are aware this is the first time tables of spin-2 spheroids  (for general complex $c$) have been presented using the AIM. Further results are presented in Sec. \ref{sec:5-3}  along with the radial QNMs for the spin two perturbations of the Kerr black hole in Table \ref{tab:Blake56S}.

%%%%%%%%%%%%%%%%%%%%%%%%%%%%%%%%%%%%%%

%==================================================
\begin{figure}[t]
  \centering
  \includegraphics[scale=.8]{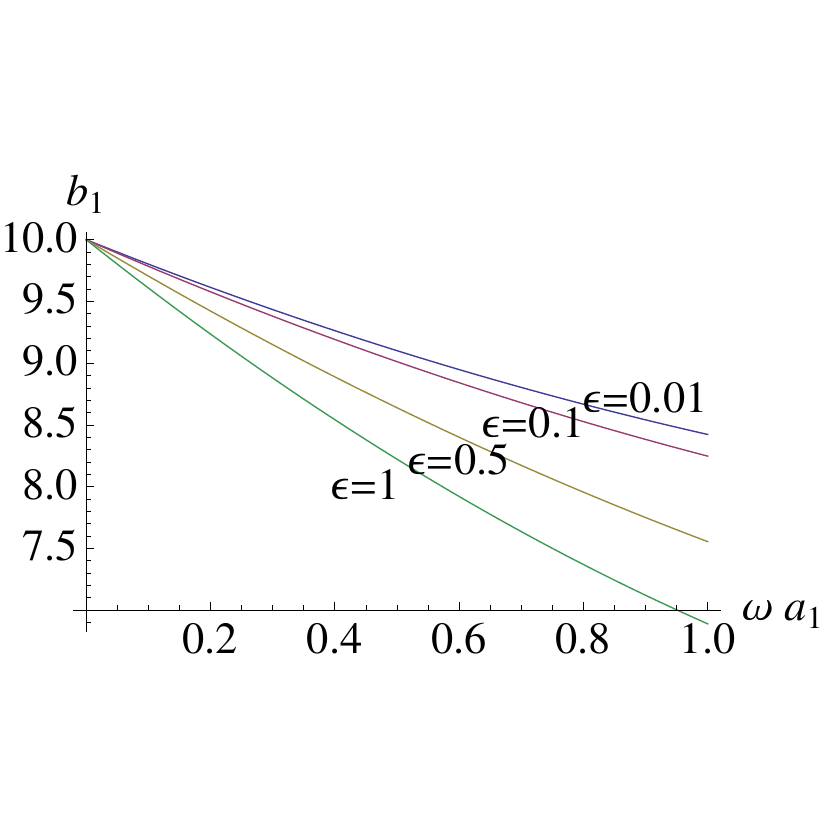}
  \includegraphics[scale=.8]{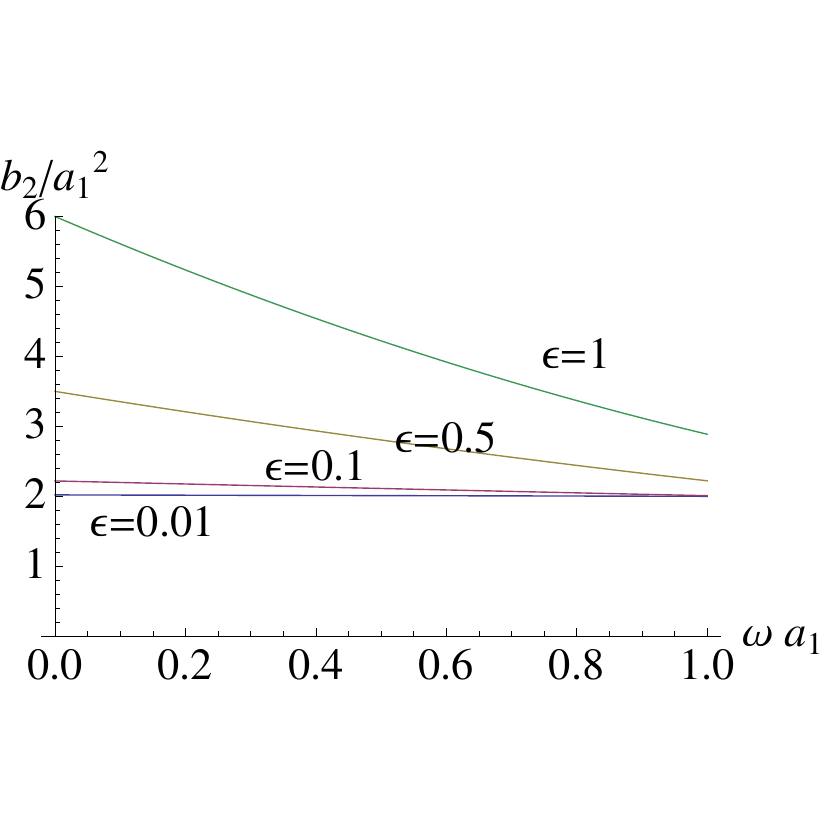}
  \caption{(Color Online) $D=6$, $g=0$, $(j,m_1,m_2,n_1,n_2)=(0,1,1,0,0)$. A plot of the eigenvalues for various choices of $\epsilon\equiv a_2/a_1$. Note that the dependence on $a_1$ has been scaled into the other quantities. \label{fig:evalepsilon}}
\end{figure}
%==================================================

%==================================================
\begin{figure}[t]
  \centering
  \includegraphics[scale=.8]{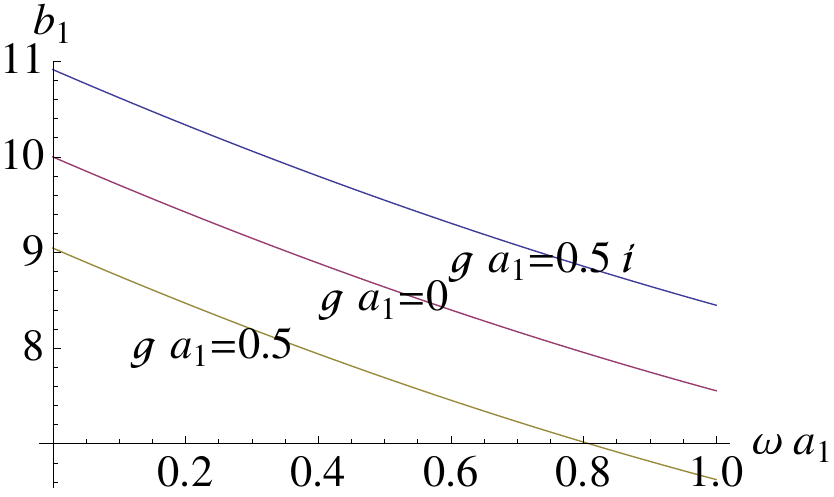}
  \includegraphics[scale=.8]{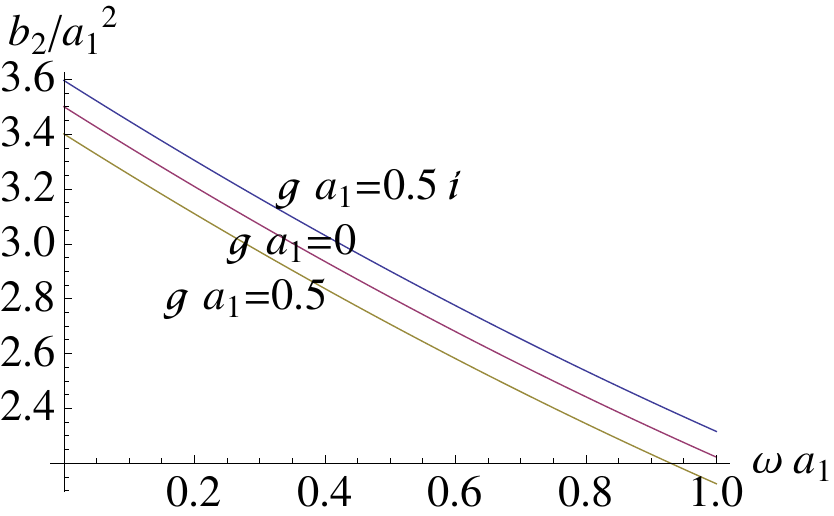}
  \caption{(Color Online) $D=6$, $\epsilon\equiv a_2/a_1=1/2$,  $(j,m_1,m_2,n_1,n_2)=(0,1,1,0,0)$. A plot of the eigenvalues for $g a_1 = 0.5 i, 0, 0.5$, corresponding to deSitter, flat, and anti-deSitter spacetimes respectively. Note that the dependence on $a_1$ has been scaled into the other quantities. \label{fig:evalg}}
\end{figure}
%==================================================

\section{Higher Dimensional Scalar Spheroidal Harmonics with two Rotation Parameters}
\label{sec:B}

\par The two Eqs. (\ref{eqn:Ryi}) are in fact the two-rotation generalization of the higher dimensional spheroidal harmonics (HSHs) studied in Ref. \cite{Berti:2005gp}. In this case, the existence of two rotation parameters leads to a system of two coupled second order 
ODEs\footnote{For the moment we are considering $\omega$ to be an independent parameter.}. In general, one would expect that the generalizations of the HSHs to $\lfloor \frac{D-1}{2}\rfloor$ rotation parameters would lead to even larger systems of equations. While these systems would also be useful generally in studies of MP black holes, here we will only focus on the two rotation case.

\par  The angular equations can be written in the Sturm-Liouville form (assuming momentarily that $\omega$ and $b_2$ are real):
\begin{equation}
  \lambda w(\xi_i) R_{\theta_i}(\xi_i)=-\frac{d}{ d\xi_i}\left(p(\xi_i) \frac{d}{d\xi_i}R_{\theta_i}(\xi_i)\right)+q(\xi_i)R_{\theta_i}(\xi_i)
\end{equation}
with the weight function $w_1(\xi_i)=\tfrac{1}{4} \xi_i^{(D-5)/2}$, the eigenvalue $\lambda=-b_1$, and 
\begin{eqnarray}
  p(\xi_i)&=&\xi_i^{(D-5)/2}\Delta_{\xi_i},\\
  q(\xi_i)&=&\frac{1}{4} \xi_i^{(D-7)/2}\left(\frac{(a_1^2-\xi_i)^2(a_2^2-\xi_i)^2}{\Delta_{\xi_i}}\tilde{\omega}_{\xi_i}^2+\frac{a_1^2 a_2^2 j(j+D-7)}{\xi_i} -b_2\right)\; ,
\end{eqnarray}
where $\Delta_{\xi_i}$ and $\tilde{\omega}_{\xi_i}$ are defined in the obvious way under the change of coordinates. Since $w(\xi)>0$ we can define the two norm's:
\begin{eqnarray}
  N_1(R_{\theta_1})&\propto&\int^{a_1^2}_{a_2^2}\xi_1^{(D-5)/2} |R_{\theta_1}|^2 d\xi_1\;,\\
  N_2(R_{\theta_2})&\propto& \int^{a_2^2}_{0}\xi_2^{(D-5)/2} |R_{\theta_2}|^2 d\xi_2\;.
\end{eqnarray}
For further details see Ref. \cite{Cho:2011yp}.

\par The regular solutions are found to be:
\begin{eqnarray}\label{eqn:modeR1}
  R_1&\sim& (\xi_1-a_2^2)^{\frac{|m_2|}{2}}(a_1^2-\xi_1)^{\frac{|m_1|}{2}}\Psi_1 \; ;\quad \xi_1\in (a_2^2,a_1^2)\; ,\\
  R_2&\sim& \xi_2^{j/2}(a_2^2-\xi_2)^{\frac{|m_2|}{2}}\Psi_2 \; ;\quad \xi_2\in (0,a_2^2) \; .\label{eqn:modeR2}
\end{eqnarray}
Now for a given value of $\omega$ we can determine $b_1$ and $b_2$ simply by performing the improved AIM \cite{Cho:2009cj} on both of the angular equations separately. This will result in two equations in the two unknowns $b_1,b_2$ which we can then be solved using a numerical routine such as the built-in Mathematica functions {\tt NSolve} or {\tt FindRoot}. More specifically we rewrite Eqs. (\ref{eqn:Ryi}) using (\ref{eqn:modeR1}) and (\ref{eqn:modeR2}) and transform them into the AIM form:
\begin{eqnarray}
  \frac{d^2\Psi_1}{d \xi_1 ^2}&=&\lambda_{01} \frac{d\Psi_1}{d \xi_1}
+s_{01} \Psi_{1}\;,\\
  \frac{d^2\Psi_2}{d \xi_2 ^2}&=&\lambda_{02} \frac{d\Psi_2}{d \xi_2}
+s_{02} \Psi_{2}\;.
\end{eqnarray} 
The AIM requires that a special point be taken about which the $\lambda_{0i}$ and $s_{0i}$ coefficients are expanded. As was shown in Ref. \cite{Cho:2009wf} different choices of this point can worsen or improve the speed of the convergence. In the absence of a clear selection criterion we simply choose this point conveniently in the middle of the domains:
\begin{eqnarray}
  \xi_{01}=\frac{a_1^2+a_2^2}{2},\quad
  \xi_{02}=\frac{a_2^2}{2}\;.
\end{eqnarray}

\par Some results are plotted in Figs. \ref{fig:evalepsilon}, \ref{fig:evalg} above. This method can now be used in the radial QNM equation in Sec. \ref{sec:6-1}.

%%%%%%%%%%%%%%%%%%%%%%%%%%%%%%%%%
%
% References
%

\end{document}